\documentclass[final,5p,times]{elsarticle}

\usepackage{graphicx}
\usepackage{amssymb}
\usepackage{amsthm}
\usepackage{lipsum}
\usepackage{xcolor}
\usepackage{comment}
\usepackage{mathtools}
\usepackage{gensymb}
\usepackage{floatrow}
\usepackage{dblfloatfix}
\usepackage{graphicx}
\usepackage{multicol} 
\usepackage{commath}
\usepackage{framed}
\usepackage{nomencl}
\usepackage{siunitx}
\makenomenclature

\renewcommand*\nompreamble{\begin{multicols}{2}}
\renewcommand*\nompostamble{\end{multicols}}

\usepackage[label font=normal]{subfig}
\usepackage{caption}
\usepackage[ruled,vlined]{algorithm2e}
\SetKwProg{Init}{Initialization:}{}{}
\theoremstyle{definition}

\usepackage{xcolor}
\usepackage[colorlinks=true]{hyperref}
	\hypersetup{
		linkcolor=blue,
		citecolor=red,
		urlcolor=green
	}

\usepackage{pifont}
\usepackage{diagbox}
\usepackage{multirow}
\newcommand{\circled}[1]{\textcircled{\footnotesize #1}}

\begin{document}
\begin{frontmatter}
\title{\LARGE Distributed Multi-Horizon Model Predictive Control for Network of Energy Hubs}

\address[label1]{Automatic Control Laboratory, Swiss Federal Institute of Technology (ETH), Z\"{u}rich, Switzerland}
\address[label2]{Urban Energy Systems Laboratory, Swiss Federal Laboratories for Materials Science and Technology (Empa), D\"{u}bendorf, Switzerland}

\author[label1,label2]{Varsha N. Behrunani\, }
\author[label2]{Hanmin Cai\, }
\author[label2]{Philipp Heer\, }
\author[label1]{Roy S. Smith\, }
\author[label1]{John Lygeros}

\begin{abstract}
The increasing penetration of renewable energy resources has transformed the energy system from traditional hierarchical energy delivery paradigm to a distributed structure. Such development is accompanied with continuous liberalization in the energy sector, giving rise to possible energy trading among networked local energy hub. 
Joint operation of such hubs can improve energy efficiency and support the integration of renewable energy resource.  
Acknowledging peer-to-peer trading between hubs, their optimal operation within the network can maximize consumption of locally produced energy. 
However, for such complex systems involving multiple stakeholders, both computational tractability and privacy concerns need to be accounted for. We investigate both decentralized and centralized model predictive control (MPC) approaches for a network of energy hubs. While the centralized control strategy offers superior performance to the decentralized method, its implementation is computationally prohibitive and raises privacy concerns, as the information of each hub has to be shared extensively. On the other hand,  a classical decentralized control approach can ease the implementation at the expense of sub-optimal performance of the overall system. In this work, a distributed scheme based on a consensus alternating direction method of multipliers (ADMM) algorithm is proposed. It combines the performance of the centralized approach with the privacy preservation of decentralized approach. A novel multi-horizon MPC framework is also introduced to increase the prediction horizon without compromising the time discretization or making the problem computationally intractable. A benchmark three-hub network is used to compare the performance of the mentioned methods. The results show superior performance of the distributed multi-horizon MPC in terms of total cost, computational time, robustness to demand and prices variations.

\end{abstract}
\end{frontmatter}
\section{Introduction}

The growth of energy demand in the last few decades coupled with climate change have resulted in an increase in environmental concerns. This has led to diversification and expansion of the technologies used to harvest and manage energy including an increased penetration of renewable energy sources in the power supply and the development of efficient multi-generation systems that promise greater energy autonomy and improved sustainability. The transition towards sustainable multi-energy systems requires the joint coordination of different interconnected energy resources and loads, as opposed to the independent planning and operation per sector (e.g., electricity, gas, etc.) of the \emph{status quo}. Furthermore, installation of technologies, such as PV, heat pumps, energy storage, etc. in residential and industrial buildings has led to the proliferation of prosumers, in the sense of units that some times act as energy consumers and others as energy producers. This has motivated research into a shift from the current grid-centric paradigm to a decentralised customer-centric topology comprising a network of multi-energy hubs.



The concept of Energy hubs was first introduced in ~\cite{Geidl:2006a}~\cite{Geidl:2007} to effectively incorporate local generation, storage and network technologies of multiple energy carriers into a unified local energy system. Energy hubs dispatch energy resources to efficiently manage time-varying production/consumption mismatches and act as intermediaries between supply and demand for a more flexible local use of energy. The coordinated operation of all the resources at the disposal of the energy hub holds leads to versatile energy management strategies and facilitates the integration of intermittent renewable energy sources. Realising this potential requires the development of advanced  control strategies for cost reduction, improving self-consumption, mitigating the adverse effects of uncertainities in demand, generation and prices, etc. 

Connecting energy hubs in a network can unlock further benefits by exploiting the link between them for peer-to-peer energy trading. In addition to lower cost and higher self-reliance, this also reduces the need for extending large-scale infrastructure, lowers the stress on the electricity grid and reduces energy imports. Harnessing these benefits requires the joint operation of hubs and a coordinated optimal dispatch of the various energy sources present in the different hubs. One approach for doing this is centralized control, where a central controller directly communicates with and controls all hubs. Many of the control approaches designed for single energy hubs such as optimal power dispatch in~\cite{almassalkhi:2011z,geng:2020r} and model predictive control (MPC) in~\cite{arnold:2009h,chandan:2014j} can be applied to the centralised control of multiple hubs~\cite{Smith:2022}. The coordinated dispatch of energy hubs with energy trading among them was first formulated as a nonlinear power flow problem in~\cite{Geidl:2006a,Geidl:2007a} and further extended in~\cite{Geidl:2007} to include storage devices. In ~\cite{Negenborn:2009,Arnold:2010}, the control is implemented using MPC considering storage systems. The centralized optimization can also use a multi-objective cost function to balance economic, environmental and social benefit costs~\cite{Scala:2014}, \cite{Maroufmashat:2015}. Such problems may be non-convex, nonlinear and non-smooth leading to computation and scalability issues~\cite{Smith:2022}. Different algorithms such as general heuristic algorithms ~\cite{Hajimiragha:2007} and genetic algorithms \cite{Eladl:2019}~\cite{Aghtaie:2014} have been proposed to mitigate these issues. Stochastic optimization methods such as the scenario approach have been proposed to account for uncertainties~\cite{Oskouei:2021} \cite{Dini:2019}. Furthermore, this strategy can also be used to effectively implement demand response programs~\cite{Yang:2016,Lu:2020}.\par
While the centralized control approach may be able to achi- eve a global optimum for the entire system, it comes with scalability and privacy concerns. As the optimization problem is non-convex and can be large, it may be prohibitive to solve in real time. Moreover, the central controller needs to collect detailed information about the demand characteristics and the converter capacities in each hub. this is information that the hub operators may not want to divulge due to privacy concerns. 

Distributed control allows the hubs to preserve privacy to an extent and the problem of scaling is less severe as much of the computation is parallelized. For this purpose, Lagrangian relaxation has been proposed for an energy hub network in \cite{Arnold:2008} and used along with MPC in \cite{Negenborn:2010}. In \cite{Xu:2018}, Lagrangian relaxation is used with stochastic optimization methods to account for  uncertainties of renewable generation and weather. The alternating direction method of multipliers (ADMM) is used in ~\cite{Nikmehr:2019}~\cite{Wang:2020} for energy systems and in conjunction with a cooperative game in \cite{Chen:2018}~\cite{Fan:2018} for the economic interaction of energy hubs in the energy sharing market for real time dispatch in \cite{Zhang:2019}. Several privacy-preserving algorithms have been used for energy hub coordination such as Benders decomposition in~\cite{Oskouei:2020}, Douglas Rachford splitting in \cite{Halvgaard:2016} and iterative mixed-integer second-order cone programming in~\cite{li:2019n}. Adopting a game theory perspective, a two-level Stackelberg game model is proposed in \cite{Wei:2017} for analyzing the multiple energy trading problem; similar bi-level formulations are used in~\cite{Huang:2020} and~\cite{Nasiri:2020}. In ~\cite{Bostan:2020}, a stochastic Stackelberg game algorithm is used that models uncertainty in electricity market price and incorporates a demand response program. Game based formulation for distributed optimization also include potential games~\cite{Bahrami:2018}, \cite{Khorasany:2021}. 

In this paper, we propose a distributed MPC control strategy based on consensus ADMM to determine the optimal operation and dispatch for a network of energy hubs. Consensus ADMM is a modification of classical ADMM in which multiple agents have to come to a consensus on a shared resource. In this method, only the value that the agents need to agree on, in this case the traded energy between the hubs, has to be communicated, improving both privacy and scalability. In addition to electrical energy, we also consider the trading of thermal energy among the hubs over a local heating grid; to the best of our knowledge, this has not been considered in earlier studies.


A key trade-off when applying MPC to energy systems is preview vs. tractability. Ideally, one would like to make the prediction horizon as long as possible to give more preview to the decisions; doing so, however, leads to a larger optimisation problem that is difficult to solve in real time. To address this trade-off, in ~\cite{Shin:2019} and ~\cite{Jiang:2019}, a temporal decomposition scheme is presented wherein the time domain is divided into multiple partially overlapping sub-problems solved in parallel. In move-blocking MPC \cite{Cagienard:2004}, the control input is forced to be constant over several steps in the horizon to reduce the dimensionality of the resulting optimisation problem which facilitates the use of a longer horizon or a smaller sampling time. A similar approach is used in \cite{LONGO:2011} using a low resolution of the MPC at some time steps and the fixing the input at these times using zero-order hold. Recent works have proposed the use of models of different granularities to extend the horizon without increasing the computational load \cite{Bathge:2016} \cite{Shin:2023}.

Here, we propose a multi-horizon MPC approach in which several models are used, each with a different time resolution. The time resolution increases later in the horizon. Each model predicts system responses for different parts of the horizon and the predictions are combined to predict system responses of the entire horizon. We exploit the structure of the energy hub and the energy systems to achieve this without adding any additional constraints. Finally, we also investigate the use and performance of the proposed method with a distributed control algorithm which has not been covered before. 

In summary, our contributions are:
\begin{enumerate}[(i)]
  \item  We formulate the model of the energy hub network propose a distributed control strategy based on Consensus ADMM that uses minimal communication between the hubs to achieve the optimal operation, mitigating concerns over privacy and scalability.
  \item We introduce a multi-horizon MPC technique to extend the prediction horizon without significantly increasing the dimension of the problem. 
\end{enumerate}

We illustrate and validate the proposed methods via extensive numerical simulations on a three-hub network, using realistic models of energy hubs and demand data. The performance of the control methods is compared in terms of cost, computation time, and scalability.

In Section II, the problem formulation and the mathematical model of the multi-carrier system are presented. In Section III, we discuss the decentralised and centralized MPC approaches and propose a distributed MPC approach based on consensus ADMM. We also develop a multi-horizon heuristic used in conjunction with the proposed control schemes. A numerical case study and simulation results applying the method to a three-hub benchmark system are presented in Section IV and V respectively. Section VI concludes this paper and outlines directions for future research.


\section{Problem Formulation and System Modelling}\label{sec:problem_statement}
 

\subsection{System Description}

\begin{figure}[t]
\centering
  \includegraphics[width=\textwidth]{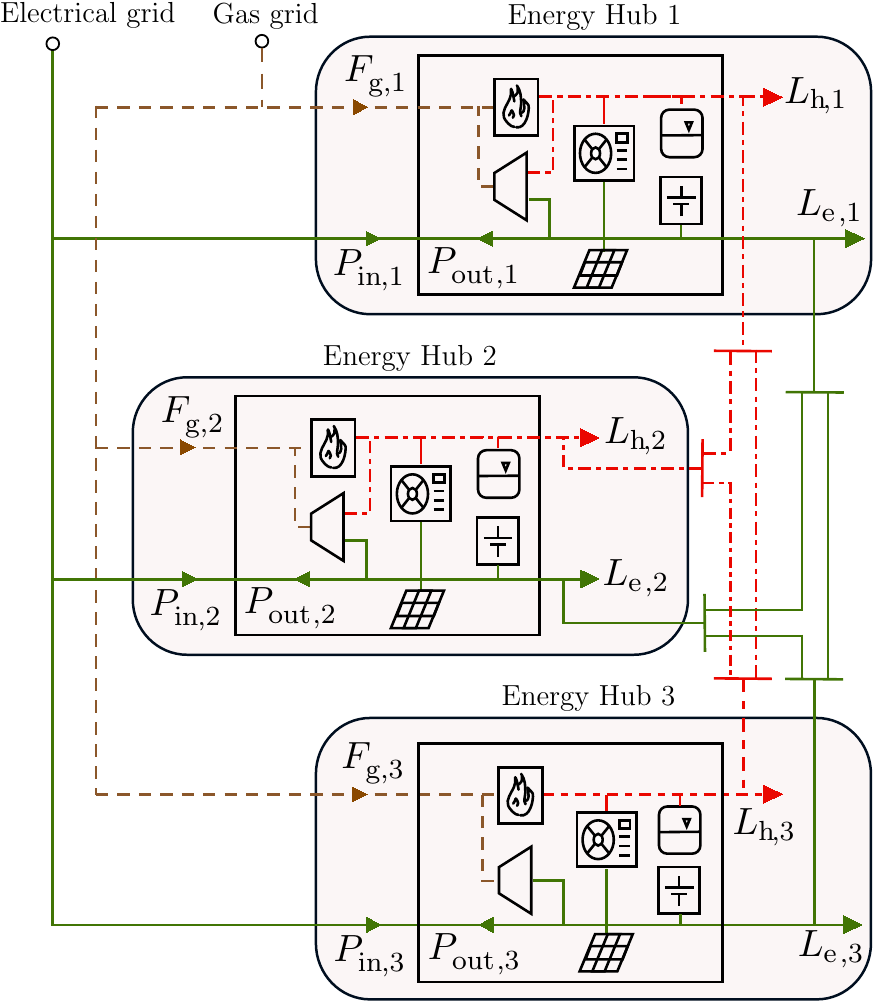}
  \caption{System of three interconnected energy hubs. Each hub can import energy from the electricity and gas grid, feed-in electricity to the grid as well as trade electrical and thermal energy with other hubs. the electricity, heating and gas network is shown in green, red and brown respectively. ~\cite{Arnold2010}.}
  \label{fig:setup_network}
 \end{figure}

We consider a general system of $H$ interconnected energy hubs labeled by $i \in \mathcal{H}:=\{1,\ldots,H\}$. To fix ideas, we use throughout a benchmark system of $H = 3$ hubs, shown in Fig. \ref{fig:setup_network} as a running example. Each hub is connected to the electricity and natural gas grid, and can trade electrical energy and thermal energy with other hubs via the electricity grid and a thermal grid, respectively. 

Each hub in the system is a general consumer equipped with energy conversion and storage devices that use electricity and gas from the grid to serve an aggregate respective electricity and heating load demand. We assume that the demands are uncontrolled, and treat demand as a disturbance from the point of view of the hub controller. We assume that the hubs contain converters such as gas boilers (GB) and heat pumps (HP), along with thermal energy storage (TS) that they can use to serve the heating demand, as well as have access to local electrical energy production using photovoltaic (PV) and electrical storage (ES) using batteries that fulfil the electricity requirement. Similarly, we assume that the hubs have converters such as solar thermal collectors (ST), Combined Heat and Power (CHP) and micro-CHP (mCHP) that simultaneously generate both electricity and heat. These devices along with the heat pump, couple the two energy systems. Electricity can also be directly drawn from the electricity grid and excess electrical energy produced in the hub can be fed back into the grid. The hubs can trade electrical energy through the existing electricity grid and are connected via a local heat distribution network that facilitates the transfer of heat energy between them.



\subsection{Energy Hub Model}

MPC makes use of models of the devices to predict the evolution of the energy hub over a finite horizon into the future. We consider discrete time models and use the superscript $k$ to denote the values of quantities at time step $k$; the superscript is omitted for quantities that are assumed to be constant.

\subsubsection{Energy Conversion Devices}
\textbf{\textit{Photovoltaic (PV) and solar thermal collectors (ST):}}\\
 The energy output of the solar photovoltaic system, $P^k_{\text{pv,i}}$, is given by, 
  \begin{equation}
\label{solar}
\begin{aligned}
P^k_{\text{pv,i}} &= \eta_{\text{pv,i}} \cdot I^k_{\text{solar,i}} \cdot a_{\text{pv,i}}\ ,
\end{aligned}
\end{equation}
 where $I^k_{\text{solar}}$ $[\text{kW/m}^2]$ is the solar irradiance incident to the surface , $a_{\text{pv}}$ is the total area of the panel, and $\eta_{\text{pv}}$ is the fixed efficiency. Similarly, the total electrical and thermal output, $P^k_{\text{st},i}$ and $Q^k_{\text{st},i}$, respectively, are given by, 

\begin{equation}
\label{CHP_fuel}
\begin{aligned}
P^k_{\text{st,i}} &= \eta_{\text{st,i}} \cdot I^k_{\text{solar,i}} \cdot a_{\text{st,i}} \cdot \alpha^{\text{p}}_{\text{st,i}} \ ,\\
Q^k_{\text{st,i}} &= \eta_{\text{st,i}} \cdot I^k_{\text{solar,i}} \cdot a_{\text{st,i}} \cdot \alpha^{\text{q}}_{\text{st,i}} \ ,
\end{aligned}
\end{equation}
where $a_{\text{st,i}}$ and $\eta_{\text{st,i}}$ are the surface area and efficiency of the solar collector respectively, and $\alpha^{\text{p}}_{\text{st},i}$ and $\alpha^{\text{q}}_{\text{st},i}$ are the fixed heat and electricity output shares of ST respectively.

\textbf{\textit{Heat pump (HP) and gas boiler (GB):}}\\
HP uses electricity, $P^k_{\text{hp,i}}$, to extract heat $Q^k_{\text{hp,i}}$ from the ground or air (ground source or air source) whereas the GB uses natural gas, $F^k_{\text{gb,i}}$, to generate heat, $Q^k_{\text{gb,i}}$. The relation between the input and output of the heat pump and the boiler are:
\begin{align}
Q^k_{\text{hp,i}} &= \text{COP} \cdot P^k_{\text{hp,i}}\ ,\\
Q^k_{\text{gb,i}} &= \eta_{\text{gb,i}} \cdot F^k_{\text{gb,i}}\ ,
\end{align}
where COP is the coefficient of performance for the pump and $\eta_{\text{gb,i}}$ is the boiler efficiency. In this work, a detailed model using part-load efficiencies is used for the boiler in which the efficiency of the boiler depends on the load, specifically, $\eta^{0.25}_{\text{gb,i}}$, $\eta^{0.5}_{\text{gb,i}}$, $\eta^{0.75}_{\text{gb,i}}$ and $\eta^{1}_{\text{gb,i}}$ for 0-25\%. 25-50\%, 50-75\% and 75-100\% operating load, respectively. This results in a peicewise linear relation that is implemented using binary variables as in~\cite{beaud:2021}.\\ 
\textbf{\textit{Combined heat and power (CHP) and
micro-CHP (mCHP):}}\\
The electrical and thermal output of the CHP are $P^k_{\text{chp,i}}$ and $Q^k_{\text{chp,i}}$, respectively. The CHP operation is bounded by a convex feasibility region described by a polyhedron with vertices A, B, C, and D~\cite{beaud:2021} and its corresponding electrical and thermal output, $p_{\text{A,i}}$, $p_{\text{B,i}}$, $p_{\text{C,i}}$,
$p_{\text{D,i}}$, and $q_{\text{A,i}}$, $q_{\text{B,i}}$, $q_{\text{C,i}}$, $q_{\text{D,i}}$, respectively. The outputs are modelled as a convex combination of the vertices with weights $w^k_{\text{A,i}}$, $w^k_{\text{B,i}}$, $w^k_{\text{C,i}}$, and $w^k_{\text{D,i}}$, respectively. the model of the CHP is given by,

\begin{equation}\label{eq:CHP}
\small
\begin{aligned}
    P^k_{\text{chp,i}} &=  \cdot \sum_{j \in S} w^k_{\text{j,i}} \cdot p_{\text{j,i}}\ ,  \ w^k_{\text{j,i}} \in [0,1] \ , \ S=\left\{\text{A, B, C, D}\right\},\\
    Q^k_{\text{chp,i}} &=  \sum_{j \in S} w^k_{\text{j,i}} \cdot q_{\text{j,i}}\ ,\\
    P^k_{\text{chp,i}} &= \eta_{\text{chp,i}} \cdot F^k_{\text{chp,i}} \ ,\\
    b^k_{\text{chp,i}}  &= \sum_{j \in S} w^k_\text{j,i}\ , \ b^k_{\text{chp,i}} \in \{0,1\} \ . 
\end{aligned}\normalsize
\end{equation}
where $F^k_{\text{chp,i}}$ is the total fuel consumed that depends only on the electrical output subject to the fuel efficiency, $\eta_{\text{chp,i}}$. The binary variable $b_{\text{chp,i}}$ is 1/0 if the CHP is On/Off. Additionally, safety constraints that limit the ramping up and down of the CHP, $r^{\mathrm{u}}_{\mathrm{chp},i}$ and $r^{\mathrm{d}}_{\mathrm{chp},i}$, and the minimum on and off time, $t^{\mathrm{u}}_{\mathrm{chp},i}$ and $t^{\mathrm{d}}_{\mathrm{chp},i}$, are also implemented using the binary variable. 
mCHP is a smaller CHP which is modelled by a simplified output model using the fixed heat and electricity output shares, $\alpha^{\text{p}}_{\text{mchp},i}$ and $\alpha^{\text{q}}_{\text{mchp},i}$, resp. and the fuel efficiency, $\eta_{\text{mchp,i}}$ by,

\begin{equation}
\label{mchp}
\begin{aligned}
P^k_{\text{mchp,i}} &= P^{\text{out,}k}_{\text{mchp,i}} \cdot \alpha^{\text{p}}_{\text{mchp,i}}\\
Q^k_{\text{mchp,i}} &=   P^{\text{out,}k}_{\text{mchp,i}} \cdot \alpha^{\text{q}}_{\text{mchp,i}}\\
P^k_{\text{mchp,i}} &= \eta_{\text{mchp,i}} \cdot F^k_{\text{mchp,i}}
\end{aligned}
\end{equation}
Additionally, the output of all the converters is limited by the following capacity constraints:
\begin{equation}
\label{eq:cLimits}
\begin{aligned}
P^{\text{min}}_{\text{m,i}} &\leq P^k_{\text{m,i}}\leq P^{\text{max}}_{\text{m,i}} \ \ \ \text{m}\in\{\text{pv, st, chp, mchp}\}\ ,\\
Q^{\text{min}}_{\text{n.i}} &\leq Q^k_{\text{n,i}}\leq Q^{\text{max}}_{\text{n.i}} \ \ \ \text{n}\in\{\text{st, gb, hp, chp, mchp}\}\ .
\end{aligned}
\end{equation}
Finally, the total electrical and thermal output of the energy converters can be compactly written as $P^k_{\text{c,i}}$ and $Q^k_{\text{c,i}}$ respectively. 

\subsubsection{Energy Storage}
The dynamics of the storage devices, in our study hot water tanks and batteries, are described by discrete time dynamical systems with a scaler state modelling the state of charge. For thermal storage, the relation between the power charged into and discharged from storage, $Q^{k}_{\text{ch,i}}$ and $Q^{k}_{\text{dc,i}}$, respectively, and the storage level, $E^{k}_{\text{ts,i}}$ is defined by
\begin{equation}
\label{str_thermal}
\begin{aligned}
E^{k+1}_{\text{ts,i}} = &\gamma_{\text{ts,i}} \cdot E^{k}_{\text{ts,i}} +  \eta_{\text{ts,i}} \cdot Q^{k}_{\text{ch,i}} - \left(\frac{1}{\eta_{\text{ts,i}}}\right) \cdot Q^{k}_{\text{dc,i}}\ ,\\
&E^{\text{min}}_{\text{ts,i}} \leq E^{k}_{\text{ts,i}} \leq E^{\text{max}}_{\text{ts,i}}\ ,\\
&Q^{\text{min}}_{\text{m,i}} \leq Q^{k}_{\text{m,i}} \leq Q^{\text{max}}_{\text{m,i}}\ \ \ \text{m}\in\{\text{ch, dc}\}\ , 
\end{aligned}
\end{equation}
where $\gamma_{\text{h,i}}$ and $\eta_{\text{h,i}}$ are the storage efficiency and charging efficiency of the thermal storage respectively (to account for standby and cycle losses). The constraints ensure that the storage levels and the power charged/discharged from storage are within some maximum and minimum levels. Similarly, for the battery storage, we use 
\begin{equation}
\label{str_battery}
\begin{aligned}
E^{k}_{\text{es,i}} &= \gamma_{\text{e,i}} \cdot E^{k-1}_{\text{es,i}} +  \eta_{\text{es,i}} \cdot P^{k}_{\text{ch,i}} - \left(\frac{1}{\eta_{\text{es,i}}}\right) \cdot P^{k}_{\text{dc,i}}\ ,\\
&E^{\text{min}}_{\text{es,i}} \leq E^{k}_{\text{es,i}} \leq E^{\text{max}}_{\text{es,i}}\ ,\\
&P^{\text{min}}_{\text{m,i}} \leq P^{k}_{\text{m,i}} \leq P^{\text{max}}_{\text{m,i}}\ \ \ \text{m}\in\{\text{ch, dc}\}\ .
\end{aligned}
\end{equation}
where  $E^{k}_{\text{es,i}}$, is the storage level, $P^{k}_{\text{ch,i}}$ is the energy charged into the battery,  $P^{k}_{\text{dc,i}}$ is the energy discharged from the battery, and $\gamma_{\text{es,i}}$ and $\eta_{\text{es,i}}$ are the standby and cycle efficiencies, respectively.
 
\subsubsection{Network}

The network and internal connections defines the energy and mass balances equations of different energy carriers.  The total gas imported into the energy hub is:
\begin{equation}
\label{fuel}
\begin{aligned}
F^{k}_{\text{g,i}} &= F^{k}_{\text{gb,i}} + F^{k}_{\text{chp,i}}  +F^{k}_{\text{mchp,i}} \ .
\end{aligned}
\end{equation}
For each hub $i$, the load balance equations for the electrical load $L^{k}_{\text{e,i}}$ is given by
\begin{equation}
\label{gen_load_ele}
\begin{aligned}
L^{k}_{\text{e,i}} &= P^{k}_{\text{c,i}}  + \left(P^{k}_{\text{in,i}}  - P^{k}_{\text{out,i}} \right)  + \left( P^{k}_{\text{dc,i}}  - P^{k}_{\text{ch,i}} \right) \ ,
\end{aligned}
\end{equation}
where $P^{k}_{\text{in,i}}$ and $P^{k}_{\text{out,i}}$ are the electrical energy imported and fed into electricity grid respectively.  Similarly, the load balance equations for the heat load $L^{k}_{\text{h,i}}$ is given by 
\begin{equation}
\label{gen_load}
\begin{aligned}
L^{k}_{\text{h,i}} &= Q^{k}_{\text{c,i}} + \left( Q^{k}_{\text{ch,i}}  - Q^{k}_{\text{dc,i}} \right) \ .
\end{aligned}
\end{equation}
In this study,  we assume there is no global thermal grid and no thermal losses in the grid within each hub. A simplified model of the thermal dynamics within the hub is considered here and a detailed model that considers temperature constraints, hydraulics, pipe dynamics, etc. is not considered here. In the absence of a heating grid, we assume demand can be met locally exactly at all times by conversion or storage. Including these in the analysis is left as a topic of future work.



\section{Control Problem Formulation}
\label{sec:methodology}

\begin{figure}[h]
\centering
  \includegraphics[width=\textwidth]{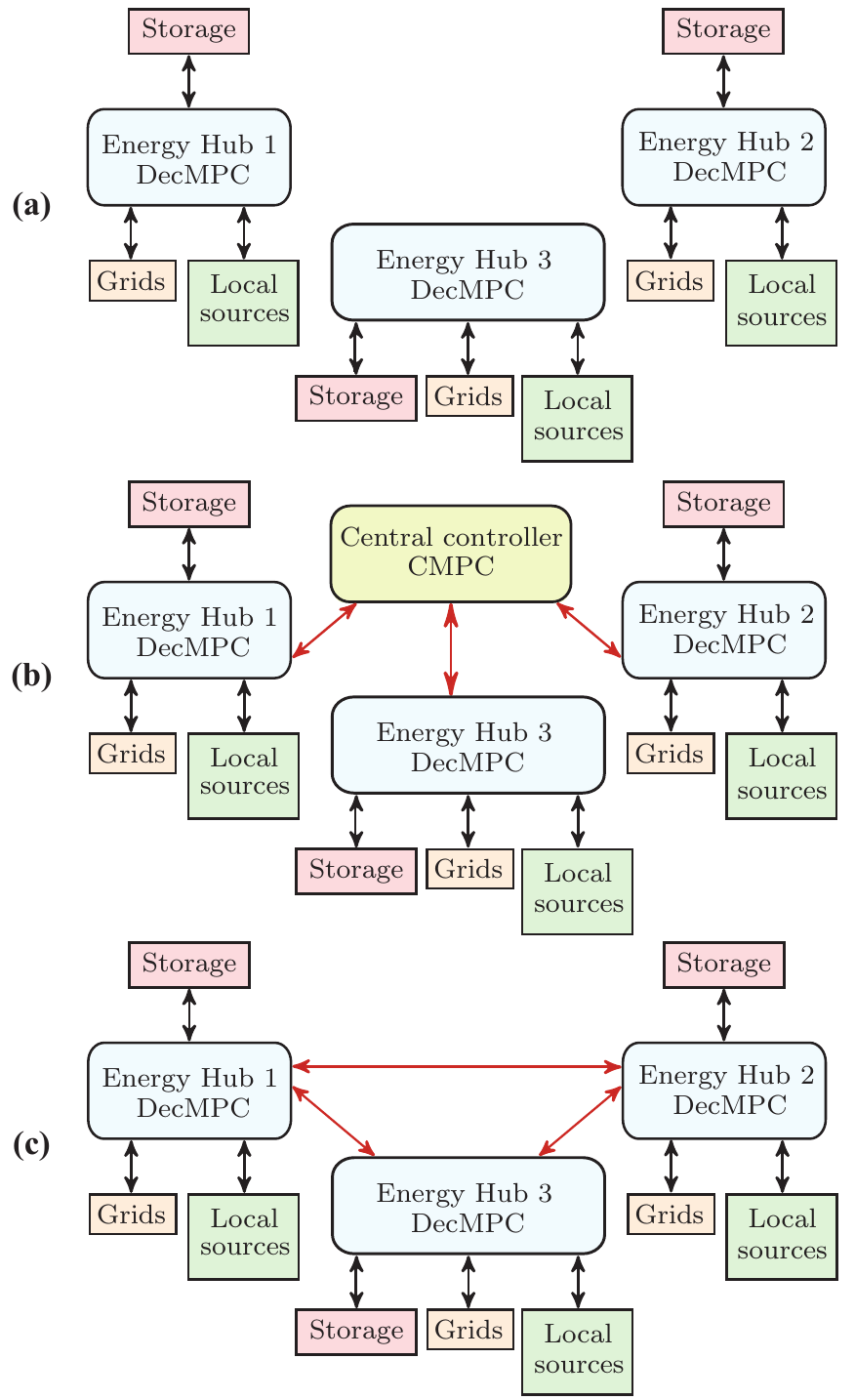}
  \caption{Different control structures: (a) Decentralised MPC (DecMPC), (b) Centralised MPC (CMPC) and (c) Distributed MPC (DMPC). The red arrows represent the communication links and the black arrows depict the physical connecitons~\cite{Smith:2022}.}
  \label{fig:control_setup}
 \end{figure}

In this section, we introduce three different control strategies for the framework described above that are illustrated in Fig. \ref{fig:control_setup}. The first two schemes are conventional, and serve as benchmarks for the novel scheme proposed in this paper. The first is a baseline decentralised MPC (DecMPC) controller where the individual hubs are operated in isolation from one another (Fig. \ref{fig:control_setup}(a)). The second is a centralised MPC (CMPC) approach with a single supervisory controller that measures all variables in the network and determines actions for all actuators (Fig. \ref{fig:control_setup}(b)). Finally, we introduce the main contribution of this work, a Distributed MPC (DMPC) method using consensus ADMM, where the controllers of the individual hubs in the network work in tandem to determine the optimal strategy by communicating limited information (Fig. \ref{fig:control_setup}(c)). 


\begin{figure}[h]
\centering
  \includegraphics[width=\textwidth]{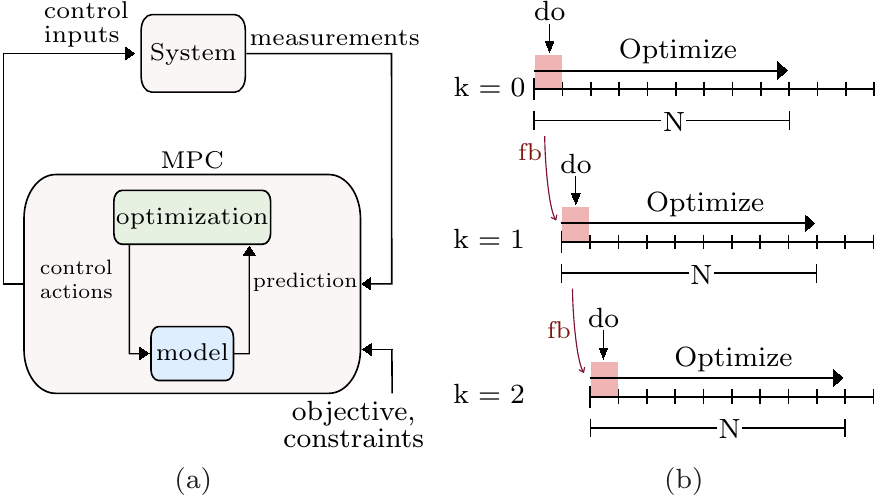}
  \caption{(a)Illustration of Model predictive control structure~\cite{Arnold2010}. (b) First three time steps of the receding horizon controller.}
  \label{fig:MPC}
 \end{figure}

All control strategies are implemented using MPC by solving iteratively a finite-horizon optimization problem involving the predicted output of a plant using its internal dynamic model. Given the model of a hub and its components, constraints, measurements at the current time, and demand forecasts, the controller formulates an open-loop optimization problem over a prediction horizon of time $T_{\text{pred}}$ divided into $N$ discrete time steps and solves it to compute a control input sequence that minimises operating costs subject to the constraints. Then, the first time step of the computed control sequence is applied to the plant~\cite{8767123}, the response is measured and the process is repeated. Fig. \ref{fig:MPC}(a) shows how MPC is implemented in closed loop. As an example, the first three time steps, $k=0,\ldots,2$, are shown in Fig. \ref{fig:MPC}(b). The receding horizon repetition brings feedback into the process through the measurements and allows the controller to continuously adapt to new forecast information, suppress the effect of model mismatch and disturbances, as well as anticipate increasing or decreasing energy prices. 

\subsection{Decentralised model predictive control (DecMPC)}
\label{decMPC}

In the decentralised MPC scheme, each hub attempts to optimize its operation individually based on its own demand and capacities, without any communication or energy exchange with the other hubs in the network. For hub $i$, let $p_{\text{c,i}} =\{p^k_{\text{c,i}}, \, p^{k+1}_{\text{c,i}}, \, \dots, p^{k+N-1}_{\text{c,i}} \}$ collect the operational set points and $p_{\text{s,i}} =\{p^k_{\text{s,i}}, \, p^{k+1}_{\text{s,i}}, \, \dots, p^{k+N-1}_{\text{s,i}} \}$  be the set of variables that are completely determined by $p_{\text{c}}$ and the constraints \eqref{solar} - \eqref{gen_load} over the horizon $k=1, \ldots, N$, where $p^{k}_{\text{s,i}}$ and $p^{k}_{\text{c,i}}$ at time step $k$ is defined as:
\begin{align}
    \nonumber p^{k}_{\text{s,i}}= & \left[E^{k}_{\text{es,i}}, E^{k}_{\text{ts,i}}, P^{k}_{\text{in,i}}, P^{k}_{\text{out,i}}\right] ^{\text{T}}\ ,\\ \nonumber
p^{k}_{\text{c,i}}=& \left[ P^{k}_{\text{pv,i}},  P^{k}_{\text{stc,i}}, P^{k}_{\text{chp,i}}, P^{k}_{\text{mchp,i}}, P^{k}_{\text{hp,i}}, P^{k}_{\text{dc,i}}, P^{k}_{\text{ch,i}},\right. \\
& \ \ \nonumber \left . Q^{k}_{\text{stc,i}}, Q^{k}_{\text{gb,i}}, Q^{k}_{\text{chp,i}}, Q^{k}_{\text{mchp,i}}, Q^{k}_{\text{hp,i}}, F^{k}_{\text{mchp,i}}, F^{k}_{\text{chp,i}}, F^{k}_{\text{gb,i}}\right]^{\mathrm{T}}\ .
\end{align}
The control objective is to minimize its total energy cost which comprises of the cost of energy procured from the electricity and gas grid.  
The cost function for hub $i$ is then
\begin{equation}
\nonumber
    J_{\text{dec,i}} \left(p_{\text{c,i}}, p_{\text{s,i}}\right) = \sum\limits_{k = 0}^{N-1} \left(c^{k}_{\text{in,e}} \cdot P^{k}_{\text{in,i}}  - c^{k}_{\text{out,e}} \cdot P^{k}_{\text{out,i}} + c^{k}_{\text{g}}  \cdot  F^{k}_{\text{g,i}}\right) \ ,
\end{equation}
where $c^{k}_{\text{in,e}}$, and $c^{k}_{\text{g}}$ are the known prices for the electricity and natural gas consumption, and $c^{k}_{\text{out,e}}$ is the feed-in tariff for the electricity grid. We assume that prices and the feed-in tariffs can be different at different time points, but that their values over the horizon are known perfectly at the time when the optimisation problem is solved. Extension to imperfect price forecasts are the topic of current work. 
The resulting decentralized finite-horizon economic dispatch problem for hub $i$ can be compactly stated as: 
\begin{equation}
\label{dec_std}
\begin{aligned}
\min_{p_{\text{c,i}}, p_{\text{s,i}}} & \ J_{\text{dec,i}} \left(p_{\text{c,i}}, p_{\text{s,i}}\right)
 \\ 
   \text{s.t. } & \text{Equations } \eqref{solar}-\eqref{gen_load} \quad \forall \  k=0, 1, \dots,N \ .\\
\end{aligned}
\end{equation}

The optimization problem is a mixed integer linear programming (MILP) problem, which can be solved using optimization problem solvers for MILP programming, such as sequential quadratic programming. 

\subsection{Centralised model predictive control (CMPC)}
\label{CMPC}
In the centralized MPC formulation, a single central controller determines the inputs and dispatch for the complete energy hub network. The central controller receives information from all the hubs and the units within as well as projected demand profiles. Hubs in the network can exchange electrical energy using the existing grid infrastructure and thermal energy using a local heat distribution network. The consolidated network with a central controller acts like a single macro-hub comprising smaller hubs that can interact and exchange energy with one another. 

The modified load balance constraints account for the energy exchange between hubs. Let $P^{k}_{\text{tr,ij}}$ and $Q^{k}_{\text{tr,ij}}$ be, respectively, the electrical and thermal energy transferred from hub $i$ to hub $j$ at time step $k$. The total electrical and thermal energy imported to hub $i$ from the neighbouring hubs are then $\sum\limits_{i\neq j} \zeta_{\text{e},ij} P^{k}_{\text{tr,ji}}$ and $\sum\limits_{i\neq j} \zeta_{\text{h},ij} Q^{k}_{\text{tr,ji}}$ respectively, where $\zeta_{\text{e,ij}}$ and $\zeta_{\text{h,ij}}$ are the corresponding efficiency of electrical and thermal energy transfer that accounts for the losses between the hubs. The loss of efficiency ensures that there are no cyclic energy transfers. Among the hubs; the network usage fees borne by the importing hub discussed below have a similar effect. For hub $i$, the resulting load balance constraints are:
\begin{align}
\nonumber L^{k}_{\text{e,i}} = & \ P^{k}_{\text{c,i}}  + \left(P^{k}_{\text{in,i}}  - P^{k}_{\text{out,i}} \right)  + \left( P^{k}_{\text{dc,i}}  - P^{k}_{\text{ch,i}} \right) \\ \label{gen_load_cen}
& + \left( \sum\limits_{j \in \mathcal{H}\backslash\{i\}} \zeta_{\text{e},ij} \cdot P^{k}_{\text{tr,ji}}  - \sum\limits_{j \in \mathcal{H}\backslash\{i\}} P^{k}_{\text{tr,ij}}\right)\\\nonumber
L^{k}_{\text{h,i}} = & \ Q^{k}_{\text{c,i}} + \left( Q^{k}_{\text{ch,i}}  - Q^{k}_{\text{dc,i}} \right) +\left( \sum\limits_{j \in \mathcal{H}\backslash\{i\}} \zeta_{\text{e},ij} Q^{k}_{\text{tr,ji}}  - \sum\limits_{j \in \mathcal{H}\backslash\{i\}} Q^{k}_{\text{tr,ij}}\right)
\end{align}

Furthermore, additional constraints are imposed to limit the trade between the hubs, in the form of line limits for the electrical and heat transfer are $\kappa_{e,ij}$ and  $\kappa_{h,ij}$ respectively:
\begin{equation}
\label{trasnfer}
    \begin{aligned}
    &P^{k}_{\text{tr,ij}}, P^{k}_{\text{tr,ji}} \leq \kappa_{\text{e,ij}}  \ ,  \\
    &Q^{k}_{\text{tr,ij}}, Q^{k}_{\text{tr,ji}}\leq \kappa_{\text{h,ij}} \ .
    \end{aligned}
\end{equation}
 If two hubs in the network are not connected to one another or cannot exchange energy between them, then the line limit is set to 0 allowing us to indirectly us to represent the network topology in the optimization.
 
In addition to the energy costs encoded in \eqref{dec_std}, the control objective also accounts for the fees collected by the network operator for using the grid infrastructure to exchange energy between the hubs; we assume that this cost is borne by the entity importing the energy. Let $c^{k}_{\text{tr}}$ be a known per-unit tariff for using the grid and $p_{\text{tr,ij}} =\{p^k_{\text{tr,ij}}, \, p^{k+1}_{\text{tr,ij}}, \, \dots, p^{k+N-1}_{\text{tr,ij}} \}$ collect all the transfer variables over the horizon $\mathcal{N}$ associated with the transfer between hub $i$ and hub $j$, with
\begin{align}
\nonumber
     p^{k}_{\text{tr,ij}}= & \left[P^{k}_{\text{ij}}, P^{k}_{\text{ji}}, Q^{k}_{\text{ij}}, Q^{k}_{\text{ji}}\right] ^{\text{T}}  \ .
\end{align}
 The resulting cost function is the sum of the cost of all the hubs in the network:
\begin{align}
\nonumber
 &J_{\text{cen}} \left(p_{\text{c}}, p_{\text{s}}, p_{\text{tr}} \right) = \\ \nonumber
    &\sum\limits_{\mathcal{H}} \underbrace{\sum\limits_{k = 0}^{N-1} \left(c^{k}_{\text{in,e}} \cdot P^{k}_{\text{in,i}}  - c^{k}_{\text{out,e}} \cdot P^{k}_{\text{out,i}} + c^{k}_{\text{g}}  \cdot  F^{k}_{\text{g,i}} + \sum\limits_{j \in \mathcal{H}\backslash\{i\}}  c^{k}_{\text{tr}} \cdot P^{k}_{\text{tr,ji}} \right)}_{J_{\text{cen,i}} \left(p_{\text{c,i}}, p_{\text{s,i}}, p_{\text{tr,ij}} \right)} \ .
\end{align}

Overall, the economic dispatch problem can be compactly written as the following MILP:
\begin{equation}
\label{cen_std}
 \begin{aligned}
\min_{p_{\text{c}}, p_{\text{s}}, p_{\text{tr}}} & \ J_{\text{cen}} \left(p_{\text{c}}, p_{\text{s}}, p_{\text{tr}} \right)
\\
\text{s.t. } & \text{Equations }  \eqref{solar}-\eqref{fuel}, \  
\eqref{gen_load_cen} - \eqref{trasnfer}\\
& \forall i \in \mathcal{H} \ \ \ \ \forall j \in \mathcal{H}\backslash\{i\} \ \ \ \ \forall \  k=0, 1, \dots,N \ .
\end{aligned}   
\end{equation}


%
%

\subsection{Distributed model predictive control (DMPC)}
\label{DMPC}

Similar to DecMPC, in the distributed MPC formulation, each energy hub has its own controller that is responsible for optimizing the decision variables of that hub. Unlike DecMPC, however, the individual controllers communicate with each other to come to an agreement on the energy transferred among the hubs.

The central control given in (\ref{cen_std}) is reformulated as a global consensus problem wherein the hubs have to reach agreement on the bilateral trades. The solution methodology for the distributed case proposed in this work is a modified version of ADMM geared towards the global consensus problem here referred to as Consensus ADMM. Consensus ADMM is a technique for turning additive objectives with constraint coupled variables, into separable objectives, that can be split between the different hubs. As a result, the implementation does not require a central coordinator and only requires information about the energy transfer to be shared among the hubs.


This is achieved by creating a local copy of the coupling global variable $p_{\text{tr,ij}}$, $\widehat{p}^{\text{ i}}_{\text{tr,ij}}$ and $\widehat{p}^{\text{ j}}_{\text{tr,ij}}$ for hub $i$ and $j$, respectively, thought of as their local estimate of the trade, along with additional constraint that ensures that it aligns with $p_{\text{tr,ij}}$ 
\begin{equation}
\label{consensus_const}
    \begin{aligned}
p_{\text{tr,ij}} - \widehat{p}^{\text{ i}}_{\text{tr,ij}} = 0 \\
p_{\text{tr,ij}} - \widehat{p}^{\text{ j}}_{\text{tr,ij}}  = 0
    \end{aligned}
\end{equation}

The central problem stated in (\ref{cen_std}) is reformulated with the new local variables and additional constraints \eqref{consensus_const}.
 Note that the load balancing constraint \eqref{gen_load_cen} in the optimization for each hub now uses the local estimate $\widehat{p}^{\text{ i}}_{\text{tr,ij}}$ instead of the global $p_{\text{tr,ij}}$ in order to completely decouple the constraints. Consensus ADMM for the resulting problem with the added constraints can be derived directly by forming the augmented Lagrangian given by:
\begin{align}
   \nonumber \sum\limits_{\mathcal{H}} \underbrace{\left(J_{\text{cen,i}}+ \sum_{j\in \mathcal{H}\setminus\{i\}} \left( \lambda_{\text{ij,i}}^{\text{T}}\left(\widehat{p}^{\text{ i}}_{\text{tr,ij}}-p_{\text{tr,ij}}\right) + \frac{\rho}{2}\left\|\widehat{p}^{\text{ i}}_{\text{tr,ij}}-p_{\text{tr,ij}}\right\|_2^2\right)\right)}_{J_{\text{dist,i}}\left(p_{\text{c,i}}, p_{\text{s,i}}, \widehat{p}^{i}_{\text{tr,ij}} \right)}
\end{align}
where $\lambda_{\text{ij,i}} =\{\lambda^{k}_{\text{ij,i}}, \, \lambda^{k+1}_{\text{ij,i}}, \, \dots, \lambda^{k+N-1}_{\text{ij,i}} \}$ is the Lagrange dual variable for hub $i$ for the transfer between hub $i$ and hub $j$ over the horizon $\mathcal{N}$ and $\rho_{\geq 0}$ is the
augmented Lagrangian penalty parameter.
The resultant dual function and its respective objective and constraints are separable as they no longer contain a coupling constraint. The resulting optimization function is solved independently at the hub level at each
iteration $s$ of the algorithm to update the local setpoints and estimates of the bilateral trade. For hub $i$, the optimization is given below.
\begin{equation}
\label{dist_hub}
 \begin{aligned}
\min_{p_{\text{c,i}}, p_{\text{s,i}},  \widehat{p}^{\text{ i}}_{\text{tr,ij}}} & \  J_{\text{dist,i}} \left(p_{\text{c,i}}, p_{\text{s,i}}, \widehat{p}^{\text{ i}}_{\text{tr,ij}} \right)
\\
\text{s.t. } & \text{Equations }  \eqref{str_thermal}-\eqref{str_battery}, 
\ \eqref{gen_load_cen}, \ \eqref{trasnfer},\ \eqref{consensus_const} \\
& \forall j \in \mathcal{H}\backslash\{i\} \ \ \ \ \forall \  k=0, 1, \dots,N \ .
\end{aligned}   
\end{equation}


Hubs $i$ and $j$ communicate their local
estimates of the trade values that are used to update the global trade decision, $p_{\text{tr,ij}}$ by an averaging step and the dual variable through residual. 
\begin{subequations}
\begin{align}
\label{eq:averaging}
\textstyle
  p^{h+1}_{\text{tr,ij}} &= \frac{1}{2} \cdot (\widehat{p}^{i,h+1}_{\text{tr,ij}} + \widehat{p}^{j,h+1}_{\text{tr,ij}}),\\
  \label{eq:dualupdate}
\lambda^{h+1}_{\text{ij,i}} &= \lambda^{h}_{\text{ij,i}} + \rho \cdot (\widehat{p}^{i,h+1}_{\text{tr,ij}} - p^{h+1}_{\text{tr,ij}}).  
\end{align}
\end{subequations}

This process continues until the local and global values of all trades converge, namely, consensus is achieved. The resulting
algorithm is summarized in Algorithm 1

\begin{figure}[t] \label{alg:dist_admm}
\flushleft
\hrule
\smallskip
\textsc{Algorithm $1$}: Distributed Consensus ADMM
\smallskip
\hrule
\medskip
\textbf{Initialization ($\boldsymbol{h=0}$)}: $p^{0}_{\text{tr,ij}}, \lambda^{0}_{\text{ij,i}} , \lambda^{0}_{\text{ij,j}} = 0$, for all trades $(i,j)$.\\ [.5em]
\textbf{Iterate until convergence:} \\ [.3em]
\hspace*{.5em}$
\left\lfloor
\begin{array}{l l}
 & \hspace*{-1em}
\text{For all hubs $i$: }\\[.1em] 
&
\hspace*{-1em}
\left\lfloor
\begin{array}{l}
\text{1. Compute $p^{h+1}_{\text{c,i}}$, $p^{h+1}_{\text{s,i}}$, and $\widehat{p}^{\text{ i,h+1}}_{\text{tr,ij}}$:}\\ 
\hspace*{1.2em} \text{given $p^{h}_{\text{tr,ij}}$, $\lambda^{h}_{ij,i}$,  solve \eqref{dist_hub}} \\[.5em]
\text{2. Broadcast $\widehat{p}^{\text{ i,h+1}}_{\text{tr,ij}}$ to hub $j$, and}
\\ \hspace*{1.2em}
\text{receive $\widehat{p}^{\text{ j,h+1}}_{\text{tr,ij}}$ from hub $j$, $\forall j\in \mathcal{N} \setminus\{i \}$,}\\[.5em]
\text{3. Update trade } p^{h+1}_{\text{tr,ij}} \text{ as in \eqref{eq:averaging}},\\[.5em]
\text{4. Update dual variable } \lambda^{h+1}_{\text{ij,i}} \text{ as in \eqref{eq:dualupdate}},\\
\hspace*{1em}
\vspace*{-1em}
\end{array}
\right.
\\[-.5em]
\hspace*{1em}\\
& \hspace*{-1em} h \leftarrow h+1
\end{array}
\right.
$

\bigskip
\hrule
\end{figure}

The algorithm converges when the 2-norm of both the primal and dual residuals converge (with a tolerance of $\epsilon$) or if the number of iterations $h$ exceeds a maximum value $h_{\text{max}}$. If consensus is reached, the algorithm terminates and outputs the value of $p_{\text{c}}$, $p_{\text{s}}$ and $p_{\text{tr}}$.





At each iteration, the sub problems are initialized with a $p^{0}_{\text{tr,ij}}$ and $, \lambda^{0}_{\text{ij,i}}$ value. The optimization in each hub is solved locally to compute optimal local decision variables, $p_{\text{c,i}}$, $p_{\text{s,i}}$, and $\widehat{p}^{\text{ i}}_{\text{tr,ij}}$ and the locally computed estimate of the trade is communicated to the other hubs. In a practical sense, this limits the communication solely to the amount of planned energy transferred between the hubs. Each hub can then compute new $p^{h+1}_{\text{tr,ij}}$, and $\lambda^{h+1}_{\text{ij,i}}$ as well as the primal and dual residual norms to check for convergence.  The newly computed values are used for the optimization in the next iteration. This continues till the primal and dual variables converge i.e., consensus is achieved between the hubs. Updating the dual variables separately drives the variables into consensus which along with the quadratic regularization helps pull the variables toward their average value while still attempting to minimize each local objective.

The Consensus ADMM based distributed optimization strategy in this case is implemented within a MPC framework which requires the algorithm presented above to be solved at every time step $k$ and the first control input is then applied to the system. In order to speed up the convergence and utilize the previous results of the distributed algorithm, we use a warm start for the MPC, in which we use the final $p_{\text{tr,ij}}$ and $ \lambda_{\text{ij,i}}$ values from the previous time step of the MPC $k-1$ to initialise the algorithm at $h = 0$ at time step $k$.

\subsection{Multi-Horizon Model Predictive Control (MH-MPC)}

 \begin{figure*}[h]
\centering
  \includegraphics[width=\textwidth]{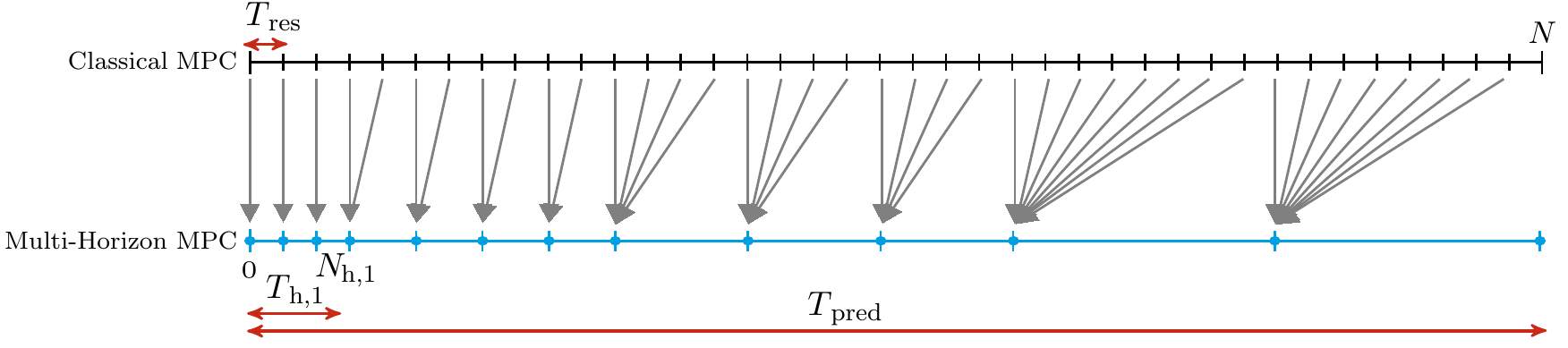}
  \caption{Comparison of Classical MPC and Multi-horizon MPC. The figure illustrates how the complete prediction horizon $T_{\text{pred}}$ is split into $N$ discrete time steps in each case and how the time steps correlate to each other. As $T_{\text{res}}$ changes over time in the multi-horizon MPC, the number of time steps at each resolution $r$ and the total time covered in each horizon are $N_{\text{h,r}}$ and $T_{\text{h,r}}$, respectively. $N_{\text{h,1}}$ and $T_{\text{h,1}}$ for the 1st time resolution is illustrated here as an example.\cite{Shin:2023}}
  \label{fig:MH-MPC}
 \end{figure*}
The control strategies described in the previous section are implemented using classical MPC. Ideally, it is desirable to have controllers with fine time discretizations and longer prediction horizons. A higher temporal resolution can capture high frequency disturbances and changes while a longer horizon can make planning decisions over a longer period of time. This is suitable in multi-energy systems which include energy carriers that operate on distinct time scales. However, increasing the horizon for a fixed temporal resolution or increasing the temporal resolution for the same prediction horizon increases the number of predicted time steps $N$ and is significantly more computationally demanding. This poses a trade-off between controller performance and computational burden. Furthermore, the uncertainty in the forecasts also increases for the far horizon which reduces the benefit of a fine resolution in the distant future. However, an approximate estimate of the future conditions may be suitable for storage planning. Currently, MPC-based solutions for energy systems are typically implemented with a short prediction horizon to reduce the impact of the load and temperature forecast uncertainties and lower the computational load. This approach, on the other hand, limits the energy-saving potential of the MPC and the optimal utilization of long-term storage.

To bridge this gap, the multi-horizon MPC framework is introduced in order to mitigate the tractability issues of MPC and extending the horizon without increasing the associated computational burden for the controller. This is done by operating at multiple horizons, each with a different time resolution which increases as we move forward in the horizon. The resulting time grid becomes more sparse as we move forward in the horizon. The difference between classical MPC and multi-horizon MPC is depicted in Fig. \ref{fig:MH-MPC}. The figure also shows the time resolution $T_{\text{res}}$, the total prediction time horizon the time resolution $T_{\text{pred}}$ and number of discrete time steps, $N$.  

This strategy relies on the property of optimal control problems known as exponential decay of sensitivity (EDS) wherein the solution sensitivity decays exponentially as one moves away from the perturbation point. In simple terms, this means that information and perturbations in the far future have a negligible impact on the current control action. The difference in the optimal solution due to any future perturbations is bounded and the solution sensitivity at a time step $i$ to any divergence or disturbance at a time step $j$ decays exponentially as the distance $|i - j|$ between time stages $i$ and $j$ increases~\cite{Shin:2023}. This result holds the key to designing multi-horizon MPC as it allows us to use the time coarsening strategies. Increasing the time resolution for the future predictions, gives an approximate prediction that does not account for the disturbances within the widening time steps. However, EDS ensures that the since these disturbances would occur in the future, their effect on the current time step decays exponentially. By having a high resolution for the first few hours that gradually increases, we ensure that the perturbations near the current time are minimized and the near-term information is captured accurately. As the horizon moves forward, while the uncertainties increase due to a lower prediction and control resolutions, these differences have a much lower effect on the optimal solution at current time step. This strategy also exploits the receding horizon nature of MPC, as only the first control action is implemented on the actual system. Any suboptimality in the future control actions will not impact the overall result as long as the first control result is unchanged.

In this work, multi-horizon MPC (MH-MPC) strategy is used with both CMPC (abbreviated as MH-CMPC) and with DMPC (abbreviated as MH-DMPC) in order to understand the effectiveness and convergence properties of the algorithms with this approach. We use an exponential time coarsening, as illustrated in Fig. \ref{fig:MH-MPC}.

It is crucial to analyse the recursive feasibility of the multi-horizon MPC scheme. In the case of the energy hub system, recursive feasibility is ensured by two aspects. The electricity grid that can be viewed as an implicit slack variable for the electrical load balance equations \eqref{gen_load} \eqref{gen_load_cen} and it can ensure that the equality constraint is met irrespective of the local generation. This allows the other generation units to produce in accordance to their own constraints and any excess/deficit can be compensated by the electrical grid. For the thermal load, in the absence of a district heating grid, the same effect is achieved in practise here by making the thermal load equality equation a soft constraint and adding a slack variable to the equation that is penalised heavily by the cost function which ensures that the equality is not violated unless needed. In practise this slack variable would result in either additional heat that would have to be discarded and deficit heat that may cause a comfort violation. Hence, these factors render feasible the overall problem at all time steps.

\section{Case Study and Numerical Simulations}\label{sec:comparisons}

In this section, a case study is presented in which the proposed distributed MPC and distributed multi-horizon MPC scheme are applied to a three-hub system. The performance of the distributed approach is compared to the performance of the central and the decentral MPC scheme. The solver Gurobi is used to perform the numerical simulations in Python. 

\subsection{Problem Configuration}

\begin{figure}[h]
\centering
  \includegraphics[width=\textwidth]{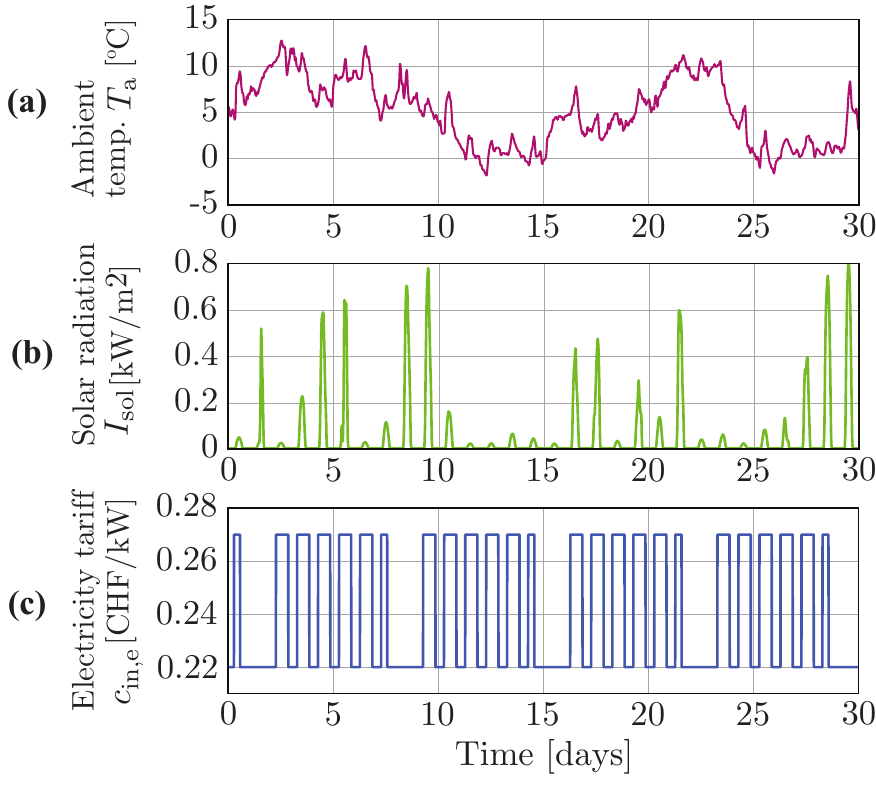}
  \caption{Inputs for the simulation for a period of 7 days. (a) Ambient temperature (b) Solar radiation (c) Electricity tariff $c_{\text{in,e}}$.}
  \label{fig:forecast}
 \end{figure}

In order to establish the efficacy of the proposed method, a sample three hub system is used. Each hub has a daily profile of its load demand and also of the energy prices that are received from the DSO. In this case study, a perfect forecast is assumed, in which no additional disturbances occur. The electrical energy prices are time varying based on the peak load hours whereas the gas prices are considered to be fixed throughout the day. The three energy hubs are interconnected as shown in Fig. \ref{fig:setup_network}.
The energy hubs are connected by a local thermal distribution grid with limited transfer capacity. The limit $\kappa_{\text{h,ij}}$ is set to $\SI{200}{\kilo\watt}$, for all connections between any hub $i$ and $j$. The energy that can be exchanged using the electrical grid is also limited and there is a fixed congestion process to use the grid. The maximum electricity that can be transferred between any 2 hubs $i$ and $j$ is $\kappa_{\text{e,ij}}$ that is set to $\SI{250}{\kilo\watt}$. Table \ref{tab: energy hub parameters} details the technologies present within the three hubs and the corresponding capacity and constraint limits for all the technologies. The tariffs for importing electricity and gas from the grid and for utilizing the grid are specified in Table \ref{tab:tariff} (the tariffs are based on the Swiss electricity prices). The system is simulated for a simulation horizon $T_{\text{sim}} = 30$ days. For specific cases, the system is also simulated for the complete year to see how the costs evolve over time.   Figure \ref{fig:forecast} shows the ambient temperature, solar radiation and electricity prices for the simulation horizon, $T_{\text{sim}}$. Finally, in order to understand how the methods scales with the number of hubs, the system is also extended up to 15 hubs.

 \begin{table}[h!]
        \centering
    \centering
    \setlength{\tabcolsep}{2pt}
    \setlength{\extrarowheight}{2pt}
    \begin{tabular}{l l l}
                \hline 
                \multicolumn{3} {c} {\textbf{Hub 1}}\\ \hline
                \multicolumn{2}{l}{Parameter} & Value \\ \hline 
                 PV &$\eta_{\text{pv},1}$, $a_{\text{pv},1}$&0.15, $\SI{8400}{\meter}^2$\\
                 &[$P_{\text{pv},1}^{\text{min}}$, $P_{\text{pv},1}^{\text{max}}$]&[0, 2500]~kW\\
                 ST &$\eta_{\text{st},1}$, $a_{\text{st},1}$, $\alpha^{\text{p}}_{\text{st},1}$, $\alpha^{\text{q}}_{\text{st},1}$ &0.15, $\SI{8400}{\meter}^2$, 0.38, 0.62\\
                 & [$P_{\text{st},1}^{\text{min}}$, $P_{\text{st},1}^{\text{max}}$] &  [0, 2500]~kW\\
                 CHP &$\eta_{\text{chp,1}}$, [$t^{\mathrm{u}}_{\mathrm{chp},1}$, $t^{\mathrm{d}}_{\mathrm{chp},1}$] &0.364, [16,4]~h\\
                & [$p_\text{A,1}$, $p_\text{B,1}$,$p_\text{C,1}$,$p_\text{D,1}$] &
              [380, 315, 745, 800]~kW 
                \\
                 & [$q_\text{A,1}$, $q_\text{B,1}$,$q_\text{C,1}$,$q_\text{D,1}$]&[0, 515, 1220, 0]~kW\\
				& [$r^{\mathrm{u}}_{\mathrm{chp},i}$, $r^{\mathrm{d}}_{\mathrm{chp},i}$] & [400,400]~kW\\
				mCHP &$\eta_{\text{mchp,1}}$, $\alpha^{\text{p}}_{\text{mchp},1}$, $\alpha^{\text{q}}_{\text{mchp},1}$ &0.35, 0.38, 0.62\\
                & [$P_{\text{mchp},1}^{\text{min}}$, $P_{\text{mchp},1}^{\text{max}}$] &
              [0, 240]~kW 
                \\
                HP & COP, [$Q_{\text{hp},1}^{\text{min}}$, $Q_{hp,1}^{\text{max}}$] & 4.5, [0, 350]~kW\\ 
                GB  & [$\eta^{0.25}_{\text{gb},1}$, $\eta^{0.5}_{\text{gb},1}$, $\eta^{0.75}_{\text{gb},1}$, $\eta^{1}_{\text{gb},1}$] & [0.59, 0.83, 0.9, 0.82]\\
                  & [$Q_{\text{gb},1}^{\text{min}}$, $Q_{\text{gb},1}^{\text{max}}$] &  [0, 350]~kW\\
                ES  & $\eta_{\text{es},1}$, $\gamma_{\text{es},1}$, [$E_{\text{es},1}^{\text{min}}$, $E_{\text{es},1}^{\text{max}}$]&0.99, 0.999, [150, 750]~kWh\\
                & [$P_{\text{ch/dc},1}^{\text{min}}$,$P_{\text{ch/dc},1}^{\text{max}}$]&[0,200]~kW\\
                TS  & $\eta_{\text{ts},1}$, $\gamma_{\text{ts},1}$, [$E_{\text{ts},1}^{\text{min}}$, $E_{\text{ts},1}^{\text{max}}$]&0.95, 0.992, [300, 12900]~kWh\\
                & [$Q_{\text{ch/dc},1}^{\text{min}}$,$Q_{\text{ch/dc},1}^{\text{max}}$] &[0,3200]~kW \\[0.5em] \hline 
                \multicolumn{3} {c} {\textbf{Hub 2}}\\ \hline
                \multicolumn{2}{l}{Parameter}  & Value \\ \hline
                PV  &$\eta_{\text{pv},2}$, $a_{\text{pv},2}$&0.15, $\SI{3170}{\meter}^2$\\ 
                &[$P_{\text{pv},2}^{\text{min}}$, $P_{\text{pv},2}^{\text{max}}$]&[0,350]~kW\\ 
                HP & COP, [$Q_{\text{hp},2}^{\text{min}}$, $Q_{hp,2}^{\text{max}}$] & 4.5, [0,350]~kW\\
                GB  & [$\eta^{0.25}_{\text{gb},2}$, $\eta^{0.5}_{\text{gb},2}$, $\eta^{0.75}_{\text{gb},2}$, $\eta^{1}_{\text{gb},2}$] & [0.59, 0.83, 0.9, 0.82]\\
                  & [$Q_{\text{gb},2}^{\text{min}}$, $Q_{\text{gb},2}^{\text{max}}$] &  [0, 50]~kW\\
                TS  & $\eta_{\text{ts},2}$, $\gamma_{\text{ts},2}$, [$E_{\text{ts},2}^{\text{min}}$, $E_{\text{ts},2}^{\text{max}}$]&0.95,0.992,[0.36, 1.62]~kWh\\
                & [$Q_{\text{ch/dc},2}^{\text{min}}$,$Q_{\text{ch/dc},2}^{\text{max}}$] &[0,0.3]~kW \\[0.5em]
                 \hline
                \multicolumn{3} {c} {\textbf{Hub 3}}\\ \hline
                \multicolumn{2}{l}{Parameter}  & Value \\ \hline
                PV  &$\eta_{\text{pv},3}$, $a_{\text{pv},3}$&0.15, $\SI{380}{\meter}^2$\\ 
                &[$P_{\text{pv},3}^{\text{min}}$, $P_{\text{pv},3}^{\text{max}}$]&[0, 80]~kW\\ 
                HP & COP, [$Q_{\text{hp},3}^{\text{min}}$, $Q_{hp,3}^{\text{max}}$] & 4.5, [0, 50]~kW\\ [1em]
        \end{tabular}
        \caption{Parameters and capacities for energy hubs used in the numerical study.}
        \label{tab: energy hub parameters} 
\end{table}

 \begin{table}[h]
        \centering
    \centering
    \begin{tabular}{l c c}
                \hline
                Tariff & Parameter & Cost(CHF/kW) \\ \hline
                Electricity - peak/offpeak & {$c_{\text{in,e}}$} & 0.27/0.22\\
                Electricity - feed-in  &$c_{\text{out,e}}$& 0.12\\ 
                Gas &$c_{\text{g}}$ & 0.115\\
                Electricity grid &$c_{\text{tr}}$& 0.02\\
        \end{tabular}
        \caption{Tariffs for electricity and gas utility.}
                \label{tab:tariff}  
                
\end{table}

The generation costs are minimized for different prediction horizons $T_{\text{pred}}$ to understand the effect of changing prediction horizon on the cost. Prediction horizons of $\SI{12}{\hour}$, $\SI{24}{\hour}$, $\SI{36}{\hour}$, $\SI{48}{\hour}$, and $\SI{72}{\hour}$ are considered. The sampling time of the plant, i.e., the time intervals in which measurements are received from the hubs $T_{\text{plant}}$ is $\SI{15}{\minute}$. Various sampling times of the MPC controller $T_{\text{res}}$ are used; $\SI{15}{\minute}$, $\SI{30}{\minute}$ and $\SI{60}{\minute}$ to understand the effect of time coarsening and the potential benefits of having a higher resolution on the cost and dispatch. The central and distributed control schemes are also implemented with multi-horizon MPC. The prediction horizon for MH-CMPC and MH-DMPC is $\SI{48}{\hour}$ and $\SI{72}{\hour}$, and the resolution of the MPC time steps ranges from $\SI{15}{\minute}$ to $\SI{6}{\hour}$ over the complete horizon as shown in Table \ref{tab:MH-MPC for 72h} with a different horizon for each $T_{\text{res}}$. The number of time steps at each resolution $r$ and the total time covered in each horizon are $N_{\text{h,r}}$ and $T_{\text{h,r}}$, respectively. 

Table \ref{tab:resolution} shows how the number of optimization time steps scale for different resolution in case of classical MPC as compared to the MH-MPC for a total time horizon of $\SI{48}{\hour}$, and $\SI{72}{\hour}$. This illustrates how significantly the number of time steps grows if a complete horizon of $T_{\text{res}} = \SI{15}{\minute}$ is used as opposed to a multi-horizon strategy that also has a highest resolution of $\SI{15}{\minute}$ but total time steps $N$ corresponding to a standard MPC resolution of $T_{\text{res}} = \SI{60}{\minute}$. Finally, the maximum number of iterations for the distributed algorithm, $h_\text{max}$, is set to 150.

\begin{table}[h]
    \centering
    \begin{tabular}{l r c r c}
                \hline
                $T_{\text{res}}$ && $N_{\text{h}}$ && $T_{\text{h}}$ [h] \\ \hline
                $\SI{15}{\minute}$ && 4 && 1\\ 
                $\SI{30}{\minute}$ &&6&& 3\\
                $\SI{1}{\hour}$  &&8&& 8\\ 
                $\SI{2}{\hour}$ &&6 && 12\\
                $\SI{4}{\hour}$  &&6&& 24\\
                $\SI{6}{\hour}$  &&4&&24\\ \hline
                &$N$ =& 34 & $T_{\text{pred}}$ =&$\SI{72}{\hour}$
                \\
                \hline
        \end{tabular}
        \caption{Number of time steps $N_{\text{h}}$ and total time covered $T_{\text{h}}$ for each resolution in MH-MPC for a total $T_{\text{pred}}$ of \SI{72}{\hour}.}
                \label{tab:MH-MPC for 72h}  
\end{table}

\begin{table}[h]
    \centering
    \begin{tabular}{l c c c c}
                \hline
                \backslashbox{$T_{\text{pred}}$}{$T_{\text{res}}$} & 15min & 30 min & 1h & Multi-horizon \\ \hline
                $\SI{12}{\hour}$ & 48 & 24 & 12 & - \\
                $\SI{24}{\hour}$ & 96 & 48 & 24 & -\\
                $\SI{36}{\hour}$ & 144 & 72 & 36 & - \\
                $\SI{48}{\hour}$ & 192 & 96 & 48 & 30 \\
                $\SI{72}{\hour}$ & 288 & 144 & 72 & 34
        \end{tabular}
        \caption{Number of optimization time steps ($N$) for all the simulation scenarios with different $T_{\text{pred}}$ and $T_{\text{res}}$.}
                \label{tab:resolution}  
\end{table}

\section{Results and Discussion}\label{sec:comparisons}
 
\subsection{Comparison of DecMPC, CMPC and DMPC}
 
Initially, the system is simulated using the decMPC, CMPC and DMPC control strategies with each of the three controllers' $T_{\text{res}}$ and for different values of the prediction horizon, $T_{\text{pred}}$. The controllers are compared based on the total operational cost i.e., the fuel, electricity and grid utility cost incurred by applying the first control input at each time step, and the total computation time required by the controllers. When the sampling time of the controller is larger than that of the plant, the system may deviate from the forecasted demand between the controller samples and this deviation may cause a mismatch in the actual load and the load that the energy hub planned to supply. For electrical load, this deviation is compensated using the electrical grid by buying additional electricity from the grid when the demand is greater than the originally forecasted demand and feeding into the grid when the electricity production exceeds the true demand. In the absence of a global thermal grid, when the thermal energy produced is more than the requested demand, this energy is discarded or regarded as waste heat and the mismatch can be quantified as a cost by computing the cost saving that could've been achieved when thermal production exceeds the true demand. Alternately, when there is a deficit in thermal energy i.e., the heat produced is less than the demand, it results in a thermal comfort violation.

\begin{figure*}[h]
\centering
  \includegraphics[width=\textwidth]{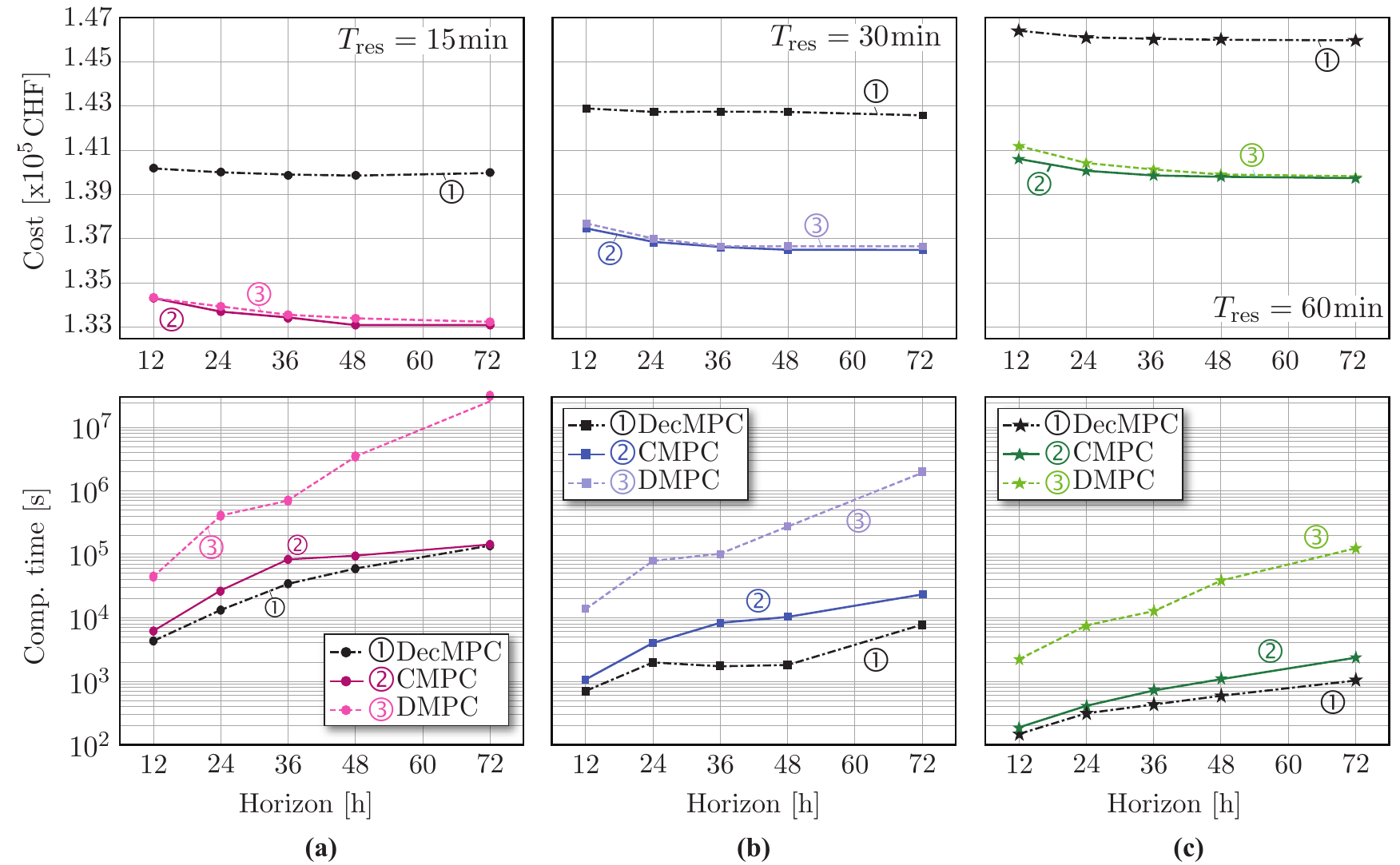}
  \caption{Comparison of cost and computation time for decMPC, CMPC and DMPC control strategies under varying $T_{\text{res}}$ and $T_{\text{pred}}$ evaluated for $T_{\text{sim}}$ = 30 days. (a),(b) and (c) show the results with $T_{\text{res}}$ of $\SI{15}{\minute}$, $\SI{30}{\minute}$, and $\SI{60}{\minute}$ respectively.}
  \label{fig:DecMPC_CMPC_DMPC}
 \end{figure*}
 
The resulting total costs and computational times over the complete simulation horizon for each of the control strategies for different $T_{\text{res}}$ and $T_{\text{pred}}$ are presented in Fig. \ref{fig:DecMPC_CMPC_DMPC}. The figure shows that irrespective of $T_{\text{res}}$ and $T_{\text{pred}}$, the central \circled{2} and distributed \circled{3} controller performance  surpasses that of the decentralised controller \circled{1} resulting in a lower cost. This is because when the hubs are operated in a coordinated manner, the controller is able to utilize the cheaper sources of energy and the storage more efficiently and trade between the hubs resulting in a lower import from the electricity grid. The distributed controller \circled{3} performs similar to the central controller \circled{2} and the solution obtained using the consensus ADMM algorithm converges to the central MPC solution. The cost of decentralisation of the DMPC can be evaluated using the optimality gap which is defined as the ratio of difference between distributed solution and the optimal central solution to the solution of the central controller. In this case, the maximum optimality gap is $\SI{0.42}{\percent}$ achieved for $T_{\text{res}} = \SI{60}{\minute}$ with $T_{\text{pred}} = \SI{12}{\hour}$, whereas the minimum is $\SI{0.07}{\percent}$ achieved for $T_{\text{res}} = \SI{15}{\minute}$ with $T_{\text{pred}} = \SI{48}{\hour}$ and the average optimality gap is $\SI{0.2}{\percent}$.

Figure \ref{fig:DecMPC_CMPC_DMPC} shows the total computational time for each of the different controller variations. The decentralised controller \circled{1} has the smallest computational time since the controllers are isolated and the resulting optimization problems are much smaller than the complete central problem. The central controller optimization problem \circled{2} combines the objectives, constraints and decision variables of each of the different hubs resulting in a much larger optimization problem to be solved at each time step. Since the optimization does not scale linearly with the number of time steps or the number of variables, the resulting time taken is much larger than just the sum of the time taken by the decentralised controllers for the three hubs. The distributed controllers \circled{3} solve an optimization problem that are similar in scale to the decentralised controller optimization problem. However, due to the iterative nature of the consensus algorithm, the optimization is solved multiple times at each time step until convergence is reached resulting in the distributed algorithm to have the largest total computational time.

Furthermore, the figure also shows how the optimal solution and total computation time is impacted by the varying prediction horizon and the controller time resolution. The total system cost for the simulation horizon decreases as the prediction horizon for the MPC controller increases and eventually, the minimum cost achievable by each controller saturates above a prediction horizon of $\SI{48}{\hour}$. This is because as the prediction horizon increases the MPC controller is able to make better decisions by anticipating the future variations but knowledge of more than $\SI{48}{\hour}$ ahead does not have a significant impact on the overall result in this case. Comparing the corresponding values from Fig. \ref{fig:DecMPC_CMPC_DMPC}(a-c), it can be seen that the sampling time of the controller can cause a significant impact on the cost with the lowest cost associated with the highest resolution.  With the increase in $T_{\text{pred}}$, the computational time for the decentral and central controller rises for all $T_{\text{res}}$ and does not scale linearly with horizon. The figure also shows that the convergence time for the distributed controller using the consensus ADMM algorithm scales exponentially with the size of the decision vector. The computation time also increases for a higher controller resolution since for the same prediction horizon $T_{\text{pred}}$, a smaller $T_{\text{res}} = \SI{15}{\minute}$ has 4 times the number of time steps($N$) as a $T_{\text{res}} = \SI{60}{\minute}$ and has to also solve the larger optimization 4 times within the same time period. This highlights the main challenge of increasing the horizon and of having a high resolution, both of which result in a much better cost optimal but causes the number of optimization steps and decision variables to rise and consequently, the time to increase exponentially justifying the need for a multi-horizon strategy.  

\begin{figure}[h]
\centering
  \includegraphics[width=\textwidth]{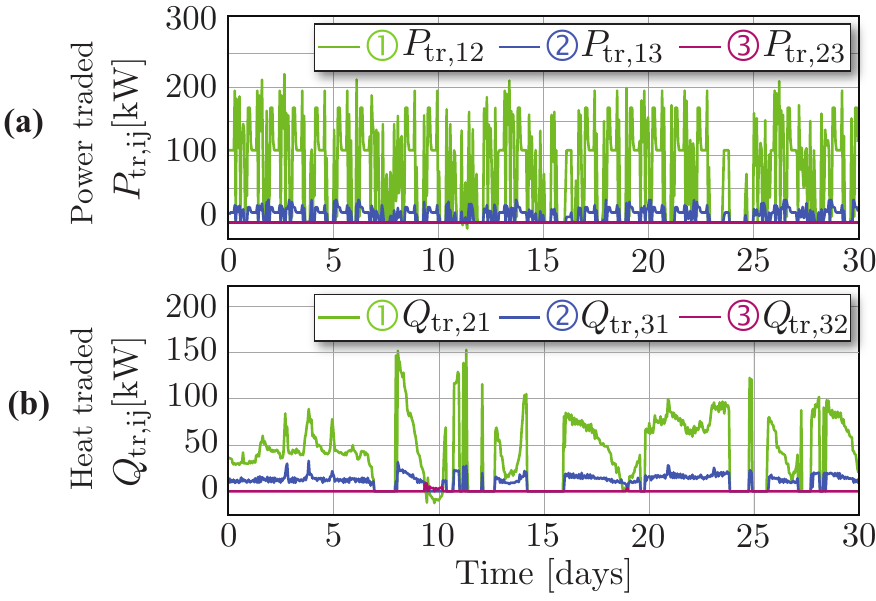}
  \caption{Total (a) Electrical energy (b) Thermal energy transferred between the three hubs over a period of 30 days. }
  \label{fig:transfer}
 \end{figure}

Figure \ref{fig:transfer} depicts the total energy transferred between the three hubs over the complete simulation. Electrical energy is traded mostly from Hub 1 to the other two hubs since it is the largest of the three hubs and has the higher production capacities (Fig.\ref{fig:transfer} (a)). This transfer is also possible due to multi-generation units such as CHP that have co-generation of electricity and heat and a much lower cost than purchasing electricity for the grid. When the heat demand is high, both electricity and heat are produced and in case of central operation, instead of exporting this excess electricity to the grid, it is transferred to neighbouring hubs to satisfy their demands. On the other hand, thermal energy is imported from the other hubs into hub 1 as seen in Fig.\ref{fig:transfer} (b). During time periods of high heating demand, the transfer of electricity allows Hubs 2 and 3 to produce more heat at a lower price using heat pumps to full capacity than if it were locally produced at Hub 1 using the more expensive gas powered boiler. This coordinated synergetic operation is what results in the lower cost for the central \circled{2} and distributed \circled{3} operation as opposed to decentralised approach \circled{1}. 


Comparing the results obtained from DMPC, CMPC and DecMPC demonstrates that the CMPC can achieve the lowest cost compared to decentralised approach, however it results in a larger optimization problem and requires the load and capacity information from all hubs. DMPC is able to achieve close to the optimal CMPC cost in a private manner using the iterative consensus ADMM algorithm. While this has a neglible optimality gap, it requires a much larger computation time due to the convergence of the iterative algorithm.

\subsection{MH-MPC - Comparison of classical MPC to MH-MPC }

In this section, the simulation results of the CMPC and the DMPC are compared to the MH-CMPC and MH-DMPC for the different $T_{\text{res}}$ and $T_{\text{pred}}$ combinations over the complete simulation.

\begin{figure}
\centering
  \includegraphics[width=\textwidth]{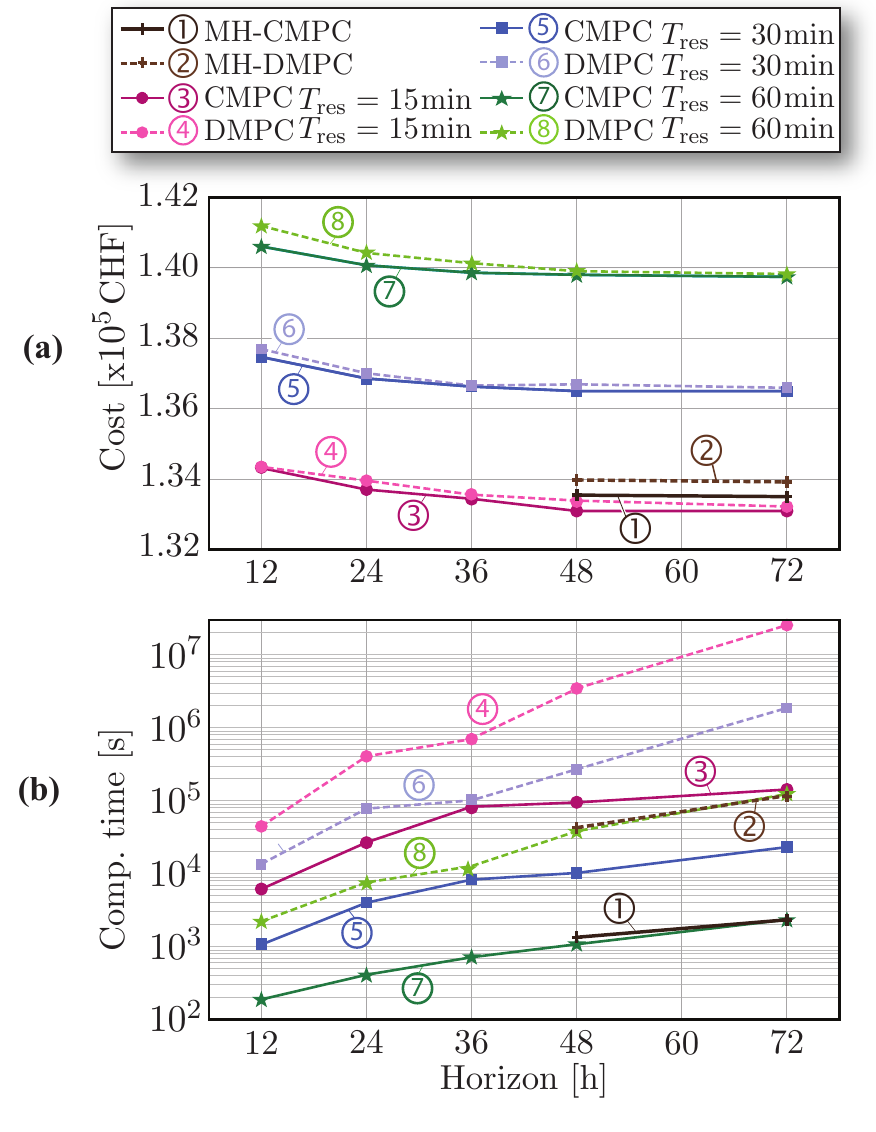}
\caption{Performance of Multi-horizon MPC controller compared to the classical MPC controller with different $T_{\text{res}}$ and $T_{\text{pred}}$ evaluated for $T_{\text{sim}}$ = 30 days. (a) Cost and (b) Total computation time of MH-MPC and standard MPC.}
  \label{fig:MHMPC_Cost}
 \end{figure}

Figure \ref{fig:MHMPC_Cost} compares the results of the MH-CMPC and MH-DMPC to the CMPC and DMPC results using different $T_{\text{res}}$ and $T_{\text{pred}}$. The cost of the MH-CMPC \circled{1} is similar to the cost of the CMPC with the same prediction horizon and the controller $T_{\text{res}}$ of $\SI{15}{\minute}$ \circled{3} despite having just a fraction of the decision variables in comparison to CMPC. Similarly, the  MH-DMPC \circled{2} also performs close to the DMPC approach with $T_{\text{res}}$ of $\SI{15}{\minute}$ \circled{4}. The optimality gap of the MH-DMPC with respect to the MH-CMPC is $\SI{0.31}{\percent}$ which is similar to the gap obtained using classical MPC. Additionally, the optimality gap of the multi-horizon approaches, MH-CMPC and MH-DMPC with respect to the classical CMPC with $T_{\text{res}} = \SI{15}{\minute}$ and $T_{\text{pred}} = \SI{72}{\hour}$ resulting in the lowest system cost over the complete simulation horizon is $\SI{0.55}{\percent}$ and $\SI{0.21}{\percent}$ respectively. This illustrates that the performance of the MH-MPC is comparable to the CMPC with a very high resolution despite having just a fraction of the decision variables in comparison to the original problem. That is because the prediction horizon of $T_{\text{pred}} = \SI{72}{\hour}$ allows MH-MPC to look far ahead into the future and find solutions compatible with the future conditions. The lower time resolution of $\SI{15}{\minute}$ in the near future makes it possible to avoid any near-term load mismatch. 
Figure \ref{fig:MHMPC_Cost}(b) shows the total computational time taken by the different control strategies. Despite having a cost comparable to the CMPC approach with $T_{\text{res}} = \SI{15}{\minute}$ \circled{3}, the MH-CMPC \circled{1} approach requires a much smaller time to compute a solution and the computation time is indistinguishable from the time taken by the CMPC with a high $T_{\text{res}} = \SI{60}{\minute}$ \circled{7} with the same $T_{\text{pred}}$. Similar to classical MPC, MH-DMPC \circled{2} in this case also takes a much larger time compared to the MH-CMPC \circled{1} since it has to solve all the subproblems repeatedly till consensus is reached. However, the total computation time of the MH-DMPC approach \circled{2} is low in comparison to other distributed controllers and matches the performance of the DMPC approach with the same $T_{\text{pred}}$ and $T_{\text{res}} = \SI{60}{\minute}$ \circled{8}. This time is negligible when compared to the DMPC with a $T_{\text{res}} = \SI{15}{\minute}$ \circled{4} despite achieving the same cost performance. While the MH-CMPC and MH-DMPC acheive optimal cost similar to the classical CMPC and DMPC with a high $T_{\text{res}} = \SI{15}{\minute}$, the computation time for MH-CMPC is approximately 60 times smaller than CMPC and that for MH-DMPC is approximately 140 times smaller than DMPC. 

While the total time taken by the MH-DMPC matches the classical DMPC at a higher $T_{\text{res}}$, it is important to note that the MH-DMPC solves 4 times as many consenses ADMM problems since the algorithm has to be applied every 15 mins as opposed to once every hour. Hence, the resulting time taken at a single time step to reach convergence and compute the MH-DMPC optimal is smaller that the time taken for DMPC. 

These results show that MH-MPC balances the trade-off between performance and computation time and achieves a low cost similar to an MPC controller with a higher sampling time and longer prediction horizon while the maintaining a low overall computation time that matches an MPC controller with a smaller sampling time.

\begin{figure}[h]
\centering
  \includegraphics[width=\textwidth]{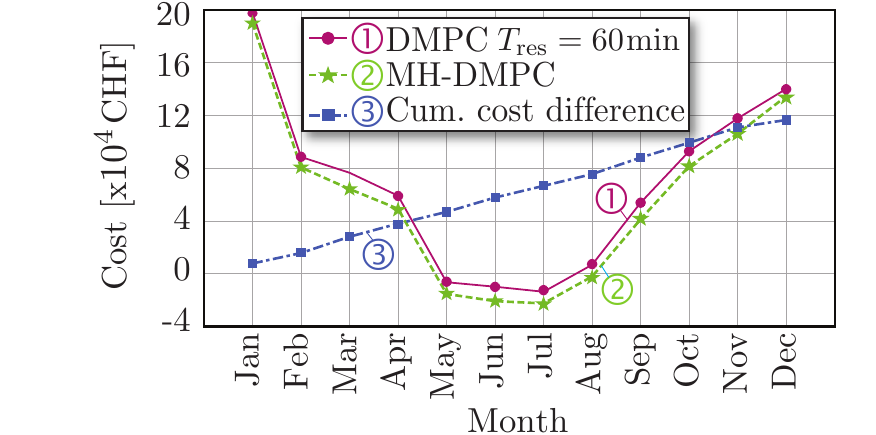}
  \caption{Cost of the MH-MPC and DMPC with$T_{\text{res}}=\SI{60}{\minute}$ and $T_{\text{pred}}=\SI{48}{\hour}$) over a period of 1 year}
  \label{fig:fullyear}
 \end{figure}

In Fig. \ref{fig:fullyear}, the performance of the MH-DMPC controller is compared to the standard DMPC with a $T_{\text{res}}$ of $\SI{60}{\minute}$, and the same $T_{\text{pred}}$ of $\SI{48}{\hour}$ for both controllers for a total $T_{\text{sim}}$ of 1 year. The results show that the MH-DMPC controller \circled{2} consistently surpasses the standard DMPC throughout the year resulting in a strictly increasing cumulative cost difference \circled{3} for every month of the year. A total cost difference amounts to more that 115 thousand Swiss francs in a single year under the current simulation setup which more than twice the average monthly cost of 57 thousand Swiss francs and higher than the actual monthly cost of 9 months of a year (Feb-Oct). While the cost difference is just a fraction of the overall cost each month and lower in the winter months compared to the summer months, the continued consistency of the MH-DMPC controller results has a significant impact over the span of just 1 year for the complete energy system. Additionally, as shown earlier, the MH-DMPC controller also results in better utility of the the renewable resources such as PV's and less energy imports from the electricity grid, therefore resulting in a higher efficiency and lower emissions for the whole system.

\begin{figure}[h]
\centering
  \includegraphics[width=\textwidth]{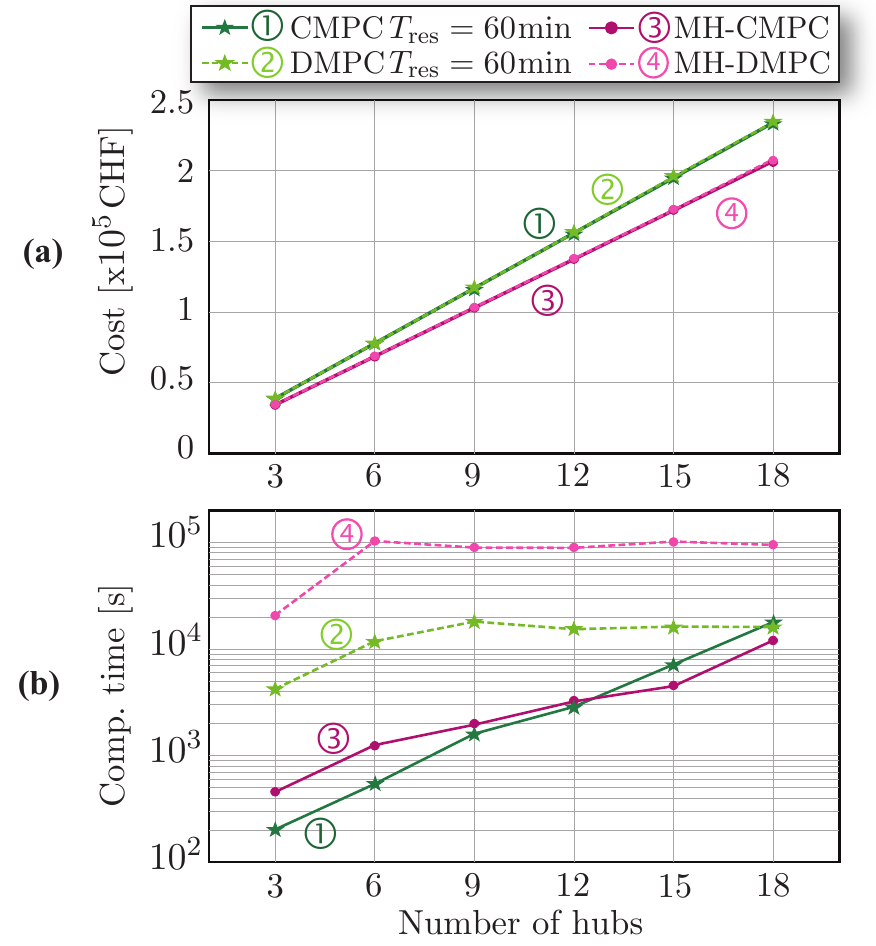}
  \caption{Results of the MH-MPC and classical MPC for different number of hubs in the network. (a)Total Cost (b) Total computation time of the MH-MPC and  DMPC controller with $T_{\text{res}}=\SI{60}{\minute}$ and  $T_{\text{pred}}=\SI{48}{\hour}$) evaluated for $T_{\text{sim}}=$ 7 days}
  \label{fig:scaling}
 \end{figure}

\begin{figure}[h]
\centering
  \includegraphics[width=\textwidth]{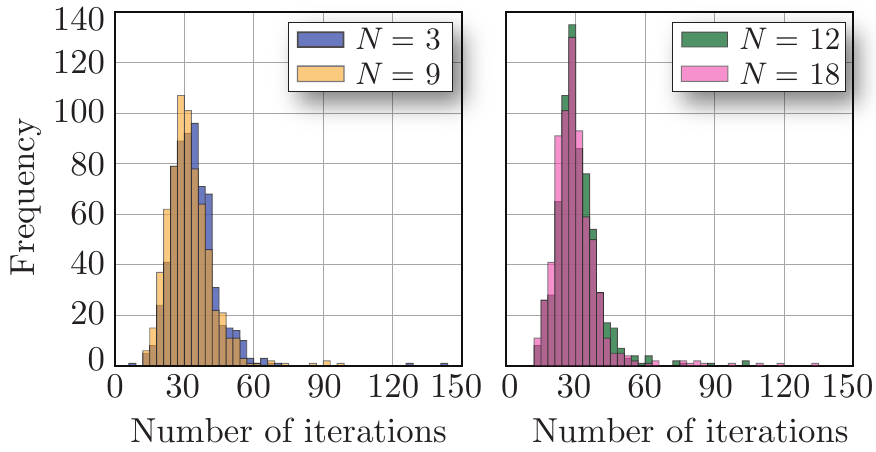}
  \caption{Histograms showing the number of iterations required for the MH-DMPC to converge and consensus to be achieved for different number of hubs in the network. (a) shows the distribution for 3 and 9 hubs and (b) shows the distribution for 12 and 18 hubs}
  \label{fig:scaling_hist}
 \end{figure}

 Finally, to understand the scalability of this approach, the system is simulated for a larger set of hubs for a period of 1 week, from 1 Dec. 2018 to 7 Dec. 2018. The results of the simulation from 3 up to 18 hubs for multi-horizon controller(MH-CMPC and MH-DMPC) and the classical MPC (CMPC and DMPC with $T_{\text{res}} = \SI{60}{\minute}$ and $T_{\text{pred}} = \SI{72}{\hour}$) are shown in Fig. \ref{fig:scaling}. In order to ensure that the distributed approaches converge within the controller time resolution even when the number of hubs in the network is large, the controller convergence criteria are modified to set a maximum time limit of $\SI{10}{\minute}$ for each time step in addition to the maximum number of iterations. Fig. \ref{fig:scaling} depicts how the total cost and required computation time of the system scales with the network. The cost of the distributed approaches are consistent with the central control methods for both the classical and MH-MPC even as the number of hubs rises. The figure verifies that performance of the MH-MPC approach surpass the classical MPC for all configurations. |
 In Fig. \ref{fig:scaling} (b), it can be seen that while the total computation times for central MPC approaches are much smaller than the distributed approaches, this time rises exponentially as the network grows. This is because both approaches solve a single optimization that becomes larger as number of hubs increases and the solution time grows exponentially. For the distributed approaches, the total time initially increases but settles as the number of hubs increases further. this is because since the computation hub level is done in parallel throughout the network, adding additional hubs does not impact the total time. Furthermore, the optimization problem of each hub remains mostly unchanged with the increase in the network size. This is further illustrated in Fig. \ref{fig:scaling_hist} where a comparison between the empirical distributions of the number of iterations required for the MH-DMPC algorithm to converge over the complete simulation for 3, 9, 12 and 18 hubs is shown. The distributions clearly verify that the number of iterations required for the distributed algorithm remains consistent irrespective of the number of hubs in the network. The mean and median number of iterations for all the simulations ranges from 28-33 the distribution is concentrated below 60, with less than 2\% of simulations requiring more than 60 iterations. This amounts to less than 13 time steps i.e., less than 3 non-consecutive hours over a week.

\section{Conclusion} \label{sec:conclusion}

The advent of multi-energy systems is transforming future systems into a network of multi-energy hubs that can produce electricity as well as trade with their peers within the network. In this paper, we present a distributed MPC approach for the coordinated operation and control of multiple energy hubs in the network.  The distributed consensus ADMM algorithm finds an optimal in a privacy-preserving manner with limited information shared between the hubs. Furthermore, a novel multi-horizon MPC scheme is employed that allows the controller to have a longer prediction horizon and a high time resolution without having a detrimental effect on the computational time. The proposed approach was tested on a simulated energy hub network. The results highlight the efficiency of the distributed approach as well as the benefit of using multi-horizon MPC in terms of total cost, required computational time, and coordination of the hubs. Future works aims to experimentally validate the proposed method on a real energy hub network, extend the method to establish fair prices for the energy traded within the network as well as reduce the need for modelling each energy hub by using data-driven methods. 
\section*{Acknowledgement}
This research is supported by the SNSF through NCCR Automation (Grant Number 180545).

\bibliographystyle{elsarticle-num-names}
\bibliography{references.bib}
\end{document}